\documentclass[tightenlines,eqsecnum,floats,aps,amsmath,amssymb%
,nofootinbib,prd,showpacs,twocolumn]{revtex4}

\usepackage{amsmath,amsthm,latexsym,amssymb,amsfonts}
\usepackage{enumerate}
\usepackage{graphicx}
\usepackage{calc}

\usepackage{LQC-symbols}

\newcommand{\spc}{\mathcal{M}}
\newcommand{\oq}{{\,{}^o\! q}}
\newcommand{\ow}{{\,{}^o\! \omega}}
\renewcommand{\oe}{{\,{}^o\! e}}
\newcommand{\cell}{\mathcal{V}}
\newcommand{\pr}{\hat{\boldsymbol{P}}}
\DeclareMathOperator*{\APS}{APS}
\DeclareMathOperator*{\sLQC}{sLQC}
\DeclareMathOperator*{\MMO}{MMO}
\DeclareMathOperator*{\B1}{B1}

\newcommand{\blA}{\boldsymbol{A}}
\newcommand{\blB}{\boldsymbol{B}}
\newcommand{\blM}{\boldsymbol{M}}
\renewcommand{\id}{\openone}

\newcommand{\Fou}{\mathcal{F}}

\begin{document}

\title{Cosmic recall and the scattering picture of Loop Quantum Cosmology}

\author{Wojciech Kami\'nski${}^{1}$}
\email{wkaminsk@fuw.edu.pl}
\author{Tomasz Paw{\l}owski${}^{2,1}$}
\email{tomasz@iem.cfmac.csic.es}

\affiliation{
  ${}^{1}$Instytut Fizyki Teoretycznej, Uniwersytet Warszawski,
  ul. Ho\.{z}a 69, 00-681 Warszawa, Poland\\
  ${}^{2}$Instituto de Estructura de la Materia, CSIC,
  Serrano 121, 28006 Madrid, Spain
}

\begin{abstract}
  The global dynamics of a homogeneous universe in Loop Quantum Cosmology
  is viewed as a scattering process of its geometrodynamical equivalent. This 
  picture is applied to build a flexible (easy to generalize) and not restricted just
  to exactly solvable models method of verifying the preservation of the 
  semiclassicality through the bounce. The devised method is next 
  applied to two simple examples: $(i)$ the isotropic Friedman Robertson Walker 
  universe, and $(ii)$ the isotropic sector of the Bianchi I model. 
  For both of them we show, that the dispersions in the logarithm of the volume $\ln(v)$ 
  and scalar field momentum $\ln(p_{\phi})$ in the distant future and past are related via 
  strong triangle inequalities. This implies in particular a strict preservation 
  of the semiclassicality (in considered degrees of freedom) in both the cases $(i)$ 
  and $(ii)$. Derived inequalities  are general: valid for all the physical states 
  within the considered models.
\end{abstract}

\pacs{04.60.Pp,04.60.Kz,98.80.Qc}
\maketitle

\section{Introduction}

Loop Quantum Gravity \cite{lqg1,lqg2} and its symmetry reduced analog, known as Loop 
Quantum Cosmology \cite{lqc1,lqc2} have experienced over recent years a dynamical progress. 
In particular, an application of the latter to the studies of a simplest (isotropic) 
models of an early Universe have shown that the quantum nature of the geometry 
qualitatively modifies the global picture of its evolution. Namely, the big bang 
singularity is dynamically resolved as it is replaced by a so called big bounce 
\cite{aps-prl} which connects the current (expanding) Universe with a contracting 
one preceding it. The results initially obtained numerically for the flat isotropic 
model with massless scalar field \cite{aps-det,aps-imp} were shown to be general 
features of that model \cite{acs,cs} and next extended to more general matter fields 
\cite{negL,posL,infl} and topologies \cite{spher,hiperb} as well as to less symmetric 
systems: anisotropic \cite{boj-b1,chiou-b1,szulc-b1,b1-mmp,b1-evo,awe,b2} and some 
classes of the inhomogeneous ones \cite{gowdy}.
Also, although the theory originates from the canonical framework, a connection 
with the \emph{spin foam} \cite{sf} models was made through the studies of the path 
integral in LQC \cite{ah} as well as the analysis of the cosmological models within 
the spin foam models themselves \cite{rv}.
Another avenue of extensions is the perturbation theory around the cosmological 
solutions \cite{pert}. The elements of its mathematical structure of LQC, which was 
initially formulated in \cite{abl}, were investigated in detail \cite{mat,kl-flat,GaSch,klp-aspects}. 
In addition to the studies performed on the genuine quantum level, there is a fast 
growing number of works employing a classical effective formulation of the dynamics 
\cite{eff-init,victor}, which often provides qualitatively new predictions \cite{eff,cs-uniq,cs-b1}. 
The methods of LQC were also applied, with various levels of rigor, outside of 
the cosmological setting, in particular in description of black hole solutions 
\cite{bh,gp-bh} and spherically symmetric spacetimes \cite{bs-spher,gp}.
There exist also studies of different prescriptions within the polymeric quantization
\cite{wigner,ma,chi} as well as of the connection between LQC and the noncommutative 
geometry \cite{bat}. The effects predicted by LQC were also applied for the regularization 
of the cosmological models not originating from the polymer quantization \cite{otros}.

The prediction of the bounce and the existence of the branch of the universe evolution 
preceding it have raised an interesting question: provided that the expanding post-bounce 
branch is semiclassical, what can we deduce about the pre-bounce one? Does it have to be 
necessarily semiclassical or can it be completely dispersed and not possible to describe 
by any classical metric? The preliminary studies performed in context of the simplification 
of LQC and presented in \cite{bojo} seemed to favor the latter possibility. However more 
detailed analysis performed in the framework of the so called \emph{solvable LQC} 
\cite{acs} have shown, that for the states satisfying quite mild semiclassicality 
assumptions for one of the branches the possible growth of the dispersion through 
the bounce is severely limited \cite{cs}. In consequence, within considered class of 
the states the semiclassicality at the distant future implied the semiclassicality in 
the distant past \cite{cs-reply}.

The latter results, although firm, strongly rely on the analytic solvability of 
the studied model, thus are difficult to extend to the original formulation of it
\cite{aps-imp} as well as to more general settings, for which one could not a priori 
exclude the loss of semiclassicality through the bounce \cite{bojo-qbb}. Finding 
a definitive answer to the question posed above requires construction of the method, 
which is sufficiently flexible to be adaptable to the situations, where the analytical 
studies fail. We introduce such method in this article, next applying it to two simple 
examples of flat models with massless scalar field as the sole matter content: $(i)$ 
an isotropic Friedman-Robertson-Walker universe, and $(ii)$ an isotropic sector of 
a Bianchi I model quantized as specified in \cite{awe}. In both cases our technique 
allows to derive certain (strong) triangle inequalities involving the dispersions 
of (the logarithms of) the total volume and the scalar field momentum. The inequalities 
are general, valid for all the physical states admitted by the model. In the case $(i)$ 
they imply an exact preservation of the semiclassicality, whereas in $(ii)$ an analogous 
result is only partial, as it does not involve all the degrees of freedom describing 
the anisotropic system.

The treatment we introduce here is based on the observation, that the structure of 
the evolution operator in LQC implies that for each physical state (in certain sense) 
there exists the geometrodynamical (Wheeler-DeWitt) one, which is its large scale 
limit in either strict \cite{aps-imp,posL} or approximate \cite{spher,negL} sense. 
Furthermore in a large class of the models the structure of the physical Hilbert 
space of each geometrodynamical theory describing the limit admits a decomposition 
onto equivalents of Klein-Gordon plane waves either incoming from or outgoing to an 
infinite volume. Since the WDW limits of LQC states are formed out of ``standing 
waves'' coupling both the incoming and outgoing ones, one can perceive the LQC 
evolution as a scattering process, which transforms the WDW incoming state (contracting 
universe) onto the outgoing one (expanding). Known properties of the limit allow 
to explicitly determine the scattering matrix and thus to relate the properties 
of the incoming and outgoing states. The procedure is explained in detail in Section 
\ref{sec:scatter} through an application to the cases $(i)$ and $(ii)$ listed above.

The paper is organized as follows: First, in Sec.~\ref{sec:frame} we briefly 
introduce the main aspects of the LQC framework used to characterize considered 
models, as well as its WDW analog. Next, in Sec.~\ref{sec:limit} we analyze in 
detail the exact WDW limit of LQC states, which is next used in Sec.~\ref{sec:scatter} 
to construct the scattering picture mentioned in the previous paragraph. That picture 
is then employed in Sec.~\ref{sec:triangle} to the analysis of the dispersions of 
the incoming/outgoing WDW states, which allows in turn to derive the triangle inequalities 
relating them. 
We also show there (Sec.~\ref{sec:tr-rel}), that the dispersions of the outgoing/incoming 
components of the WDW limit equal the dispersions of a genuine LQC state in the asymptotic 
future and past, thus extending the inequalities to these quantities.
We conclude with Section \ref{sec:concl} with the discussion of the main results as 
well as the possibilities of their extensions to more general settings. In order to 
make the presentation of the studies clearer to the reader, some details of the 
mathematical studies as well as the details of the numerical analysis used in the 
article are moved to Appendix~\ref{app:math} and \ref{app:num} respectively.

\section{The framework of LQC and its WDW analog} \label{sec:frame}

In this section we briefly sketch the specification, quantization program and relevant 
properties of the models we are going to study. We discuss only those elements of the 
theory, which are relevant for our analysis. For the detailed description of the 
quantization and the properties of the systems the reader is referred to \cite{abl,aps-imp} 
and \cite{awe}.

Since the considered systems are constrained ones, they are quantized via Dirac program, 
which consists of the following steps: kinematical quantization ignoring the constraints, 
promoting the constraints to quantum operators and finding the physical Hilbert space built
out of the states annihilated by the quantum constraint. In accordance to this procedure this 
section has the following structure: First we introduce the classical models, LQC kinematics
and quantization of the constraint for the FRW and Bianchi I models separately in 
Sec.~\ref{sec:frame-FRW} and \ref{sec:frame-B1} respectively. Next we describe the 
structure of the  physical Hilbert space and relevant observables for LQC theory 
(in Sec.~\ref{sec:frame-phys}), as well as its WDW analog (Sec.~\ref{sec:frame-WDW}).

\subsection{Isotropic flat FRW universe}\label{sec:frame-FRW}

\subsubsection{Classical theory}\label{sec:frame-FRW-class}

On the classical level spacetimes of this class admit a (parametrized by 
a time $t$) foliation by homogeneous surfaces $\spc=\Sigma\times\re$, where 
$\Sigma$ is topologically $\re^3$. Their metric tensor is 
\begin{equation}\label{eq:metric}
  g = -N^2\rd t^2 + a^2(t) \oq , 
\end{equation}
where $N$ is a lapse function, $a$ is a scale factor (or equivalently a size of 
certain selected region $\cell$, see the discussion below \eqref{eq:Cgr-def}) 
and $\oq$ is a fiducial Cartesian metric. 
To describe the system further we apply the canonical formalism, expressing the 
geometry in terms of Ashtekar variables: connections and triads, selecting the 
gauge fixing in which they can be expressed in terms of the real connection and 
triad coefficients $c,p$ 
\begin{subequations}\label{eq:Ashtekar}\begin{align}
  A^i_a &= c \, V_o^{-\frac{1}{3}} \ow^i_a  , &  
  E^a_i &= p \, V_o^{-\frac{2}{3}} \sqrt{\oq} \oe^a_i , 
\tag{\ref{eq:Ashtekar}}\end{align}\end{subequations}
where $\oe^a_i$/$\ow^i_a$ is an orthonormal triad/cotriad corresponding to the fiducial 
metric $\oq$ and $V_o$ is the fiducial volume of $\cell$. The variables $c,p$ are 
canonically conjugated with $\{c,p\}=8\pi G\gamma/3$ (where $\gamma$ is a Barbero-Immirzi 
parameter, which value has been set following \cite{gamma}) and are global degrees 
of freedom of the geometry. In particular $p=a^2$.

Within selected gauge all the constraints except the Hamiltonian one are automatically 
satisfied. The remaining constraint takes the form $C = N(C_{\rm gr}+C_{\phi})$ where
\begin{equation}\label{eq:Cgr-def}
  C_{\rm gr} = -\frac{1}{\gamma^2}\int_{\cell}\rd^3x\, \epsilon_{ijk} e^{-1} 
  E^{ai} E^{bj} F^k_{ab} ,
\end{equation}
with $e:=\sqrt{|\det(E)|}$ and $F$ being a curvature of a connection $A$. To deal 
with the noncompactness of $\Sigma$ the integration of a Hamiltonian density was 
performed only over a chosen cubic cell $\cell$ constant in comoving coordinates,
which is an equivalent to the infrared cut-off. Despite this, the physical 
predictions are invariant with respect to the choice of the cell \cite{aps-imp}.

The remaining matter part of the constraint equals
\begin{equation}\label{eq:Cphi-def}
   C_{\phi} = 8\pi G p_{\phi}^2 / p^{3/2} ,
\end{equation}
where $\phi$ and $p_{\phi}$ are, respectively, the value of the scalar field and 
its conjugate momentum with $\{\phi,p_{\phi}\}=1$.

\subsubsection{Loop quantization}\label{sec:frame-FRW-q}

The classical system specified in Sec.~\ref{sec:frame-FRW-class} is next quantized 
via methods of Loop Quantum Gravity. In particular, the Dirac program is employed 
to construct the physical Hilbert space. It consists of the following steps:

$\bullet$ \emph{Quantization on the kinematical level (ignoring the constraints).}
Here, the as the basic objects instead of $A^i_a$, $E^a_i$ we select the holonomies of
$A^i_a$ along the straight lines and fluxes of $E^a_i$ along the unit square $2$-surfaces, which
form a closed algebra. 
The direct implementation of the procedure used in LQG \cite{lqg2} leads to the 
gravitational Hilbert space $\Hil_{\gr} = L^2(\bar{\re},\rd\mu_{\Bohr})$, where 
$\bar{\re}$ is the Bohr compactification of the real line. The basic operators 
are respectively the holonomies $\hat{h}^{(\lambda)}$ along the edge of the length 
$\lambda$ and triad $\hat{p}$ (corresponding to the flux across the unit square). 
The basis of $\Hil_{\gr}$ is built of the eigenstates of $\hat{p}$ and parametrized 
by $v$ such that $\hat{p}\ket{v} = (2\pi\gamma\lPl^2\sqrt{\Delta})^{2/3} \sgn(v)
|v|^{2/3}\ket{v}$, where the parameter $\Delta$ is the so called \emph{area gap} 
specified in the next point. The inner product on $\Hil_{\gr}$ is given by
\begin{equation}\label{eq:gr-ip}
  \sip{\psi}{\chi} = \sum_{v\in\re} \bar{\psi}(v)\chi(v).
\end{equation}

The matter degrees of freedom are quantized via standard methods, thus attaining the 
Schroedinger-like representation. In consequence the full kinematical Hilbert space
takes the form 
\begin{equation}
  \Hil_{\kin} = \Hil_{\gr} \otimes \Hil_{\phi} , \quad \Hil_{\phi} := L^2(\re,\rd\phi).
\end{equation}
The basic operators on $\Hil_{\phi}$ are the field value $\hat{\phi}$ and its momentum 
$\hat{p}_{\phi} = -i\hbar\partial_{\phi}$.

$\bullet$ \emph{Promoting the constraint to quantum operator.} 
For that all the 
geometric components in \eqref{eq:Cgr-def} and \eqref{eq:Cphi-def} have to be 
expressed first in terms of the holonomies and fluxes, which is essentially done 
via methods specified in \cite{Th-trick}. The field strength $F^k_{ab}$ is in particular 
represented via holonomies along the closed square loop. The requirement for our 
theory to mimic the properties of LQG and the discreteness of the area operator 
$\hat{\rm Ar}$ there forces us to fix the physical area $\Delta$ of this loop as 
the $1$st nonzero eigenvalue of $\hat{\rm Ar}$, which is the unique physically 
consistent choice for that technique \cite{cs-uniq}. 

Presently in the literature there exists several prescriptions of constructing the quantum 
Hamiltonian constraint differing by fine details, like the lapse, the factor ordering 
and the symmetrization. Here we focus on three of them, defined in \cite{aps-imp}, 
\cite{acs} and \cite{b1-mmp,mmo} and denoted, respectively, as the \emph{APS}, \emph{sLQC} and \emph{MMO}
prescription.   
In all these cases the resulting operator can be brought to the form
\begin{equation}\label{eq:constr-quant}
  \id\otimes\partial_{\phi}^2 + \Theta \otimes \id ,
\end{equation}
where an action of the operator $\Theta$ equals
\begin{equation}\label{eq:ev2-gen}\begin{split}
  -[\Theta\psi](v) &= f_+(v)\psi(v-4) - f_o(v)\psi(v) \\ 
  &+ f_-(v)\psi(v+4) ,
\end{split}\end{equation}
with the form of $f_{o,\pm}$ depending on the particular prescription used and given 
respectively by
\begin{itemize}
  \item APS: 
    \begin{subequations}\label{eq:coeffs-first}\begin{align}
      f_\pm(v) &= [B(v\pm 4)]^{-\frac{1}{2}}\tilde{f}(v\pm 2)[B(v)]^{-\frac{1}{2}} , \\
      f_o(v) &= [B(v)]^{-1} [f_+(v)+f_-(v)] ,
    \end{align}\end{subequations}
    where \cite{rep-diff}
    \begin{subequations}\begin{align}
      \tilde{f}(v) &= (3\pi G/8) |v| \big| |v+1| - |v-1| \big| \\
      B(v) &= (27/8) |v| \big| |v+1|^{1/3} - |v-1|^{1/3} \big|^3
    \end{align}\end{subequations}
  \item sLQC: 
    \begin{subequations}\label{eq:fslqc}\begin{align}
      f_\pm(v) &= (3\pi G/4) \sqrt{v(v\pm 4)}  (v\pm 2) , \\  
      f_o(v) &= (3\pi G/2) v^2 ,
    \end{align}\end{subequations}
  \item MMO:
    \begin{subequations}\label{eq:fmmo}\begin{align}
      f_{\pm}(v) &= C g(v\pm 4)s_{\pm}(v\pm 2)g^2(v\pm 2)s_{\pm}(v)g(v), \notag \\
      f_o(v) &= C g^2(v) [ g^2(v-2)s^2_-(v) + g^2(v+2) s^2_+(v) ] , \tag{\ref{eq:fmmo}}
    \end{align}\end{subequations}
    where
    \begin{subequations}\label{eq:Theta-form-last}\begin{align}
      g(v) &= \big| |1+1/v|^{1/3} - |1-1/v|^{1/3} \big|^{-1/2} , \\
      s_{\pm}(v) &= \sgn(v\pm 2) + \sgn(v) , \\
      C &= \pi G/12 .
    \end{align}\end{subequations}
\end{itemize}
The operator $\Theta$ is denoted, respectively, by $\Theta_{\APS}$, $\Theta_{sLQC}$ 
and $\Theta_{\MMO}$ and is well defined in all the listed prescriptions in particular for 
$\varepsilon=0$ (see the detailed discussion in \cite{klp-aspects} for APS and
\cite{mmo} for MMO).

$\bullet$ \emph{Building a physical space out of states annihilated by the constraint.}
Since the operator \eqref{eq:constr-quant} is essentially self-adjoint \cite{kl-flat}, 
this step can be realized via systematic procedure of the \emph{group averaging} 
\cite{gave,gave2}. On the other hand its form selects another natural way of finding 
the solutions \cite{aps-imp,GaSch} which in this case is equivalent to it, namely
the reinterpretation of the constraint as the Klein-Gordon-like equation
\begin{equation}\label{eq:constr-KG}
  [\partial^2_\phi\Psi](v,\phi) = -[\Theta\Psi](v,\phi) ,
\end{equation}
defining the evolution of a free system along $\phi$, that is the mapping between
the the spaces of the ``initial data'' -- restrictions of $\Psi$ to the surfaces of constant
$\phi$
\begin{equation}\label{eq:evo-map}
  \re\ni\phi \mapsto \Psi(\cdot,\phi) \in \Hil_{\gr} .
\end{equation}

The quite simple form of the operator $\Theta$ allows to easily define the physical 
Hilbert space $\Hil_{\phy}$ via its spectral decomposition. This step, as well as the notion 
of evolution will be described in more detail in Sec.~\ref{sec:frame-phys}.

Before going to it let us note, that the structure of $\Theta$ and \eqref{eq:constr-quant} 
provides the natural division of the domain of $v$ onto the subsets (the \emph{lattices}) 
\begin{equation}
  \lat_{\varepsilon} = \{ \varepsilon+4n;\ n\in\integ\} ,\quad \varepsilon \in [0,4[
\end{equation}
preserved by the action of $\Theta$. This division is naturally transferred to the 
splitting of $\Hil_{\phy}$ onto the superselection sectors. In consequence it is enough 
to fix particular value of $\varepsilon$ and work just with the restriction of the domain
of $\Theta$ to functions supported on $\lat_{\varepsilon}$ only. For the clarity we 
will consider just the sector $\varepsilon=0$, however the presented treatment and 
its results generalize easily (at the qualitative level) to all the sectors.

Further simplification comes from the fact, that the considered system does not admit 
parity violating interactions. In consequence the triad orientation reflection 
$v\mapsto -v$ being the large symmetry provides another natural division onto superselection
sectors, namely the spaces of symmetric and antisymmetric states. For the selected 
sector $\varepsilon=0$ this particular choice allows to further restrict the support 
of the functions to $\lat^+_{0} := \lat^+_{0} \cap \re^+$.

\subsection{Flat Bianchi I universe}\label{sec:frame-B1}

The first step in the generalization of the model presented in previous Section 
is an extension to the flat Bianchi I model, describing the universe with the same 
matter content and topology, which however while being still homogeneous is not 
necessarily isotropic. Its (preliminary) analysis within LQC has been initiated 
in \cite{boj-b1}. Later more detailed analysis of its kinematics and dynamics was 
performed in \cite{chiou-b1} and \cite{szulc-b1} (see also \cite{b1-mmp,b1-evo} 
for a vacuum  case), although the quantization prescription used there is not 
applicable to the noncompact cases \cite{cs-b1}. The first description valid also 
in noncompact situation was constructed in \cite{awe}, which we will follow in this 
article. In this section we briefly introduce those elements of the framework, which 
are needed as a basis for our analysis. The treatment is in fact an extension 
of the one applied to the FRW universe. Therefore, for shortness, here we will
focus just on these aspects of it, which differ from the description presented in
Sec.~\ref{sec:frame-FRW}. For the detailed description the reader is referred to 
\cite{awe}.

\subsubsection{The classical model}

Classically the Bianchi I flat spacetime admits the same foliation by homogeneous 
surfaces as the FRW one. The general form of the metric is
\begin{equation}
  g = -N^2\rd t^2 + a^2_1(t)\rd {x_1}^2 + a^2_2(t)\rd {x_2}^2 + a^2_3(t)\rd {x_3}^2 
\end{equation}
where $x_i$ are chosen (comoving) coordinates defining the fiducial Cartesian metric
$\oq := \rd {x_1}^2 + \rd {x_2}^2 + \rd {x_3}^2$ with orthonormal (co)triad ($\ow^i_a$) 
$\oe^a_i$ . To construct the Hamiltonian formalism one has to introduce again 
a fiducial cell, which here is described by three fiducial (i.e. with respect to 
the metric $\oq$) lengths $L_i$ and fiducial volume $V_o:=L_1L_2L_3$. Here it is 
again possible to fix a gauge in which the Ashtekar connections and triads are 
represented by three pairs of canonically conjugated coefficients $c^i, p_i$
\begin{subequations}\label{eq:Ashtekar-B1}\begin{align}
  A^i_a &= c^i \, L_i^{-1} \ow^i_a  , &  
  E^a_i &= p_i \, L_i V_o^{-1} \sqrt{\oq} \oe^a_i , 
\tag{\ref{eq:Ashtekar-B1}}\end{align}\end{subequations}
and all the constraints except the Hamiltonian one are automatically solved. 
The Poisson brackets between coefficients equal $\{c^i,p_j\}=8\pi G\gamma\delta^i_j$.

Following \cite{awe} we choose the lapse $N=\sqrt{|p_1p_2p_3|}$.
The Hamiltonian constraint has the same form as in the FRW case, and its 
components are given by \eqref{eq:Cgr-def} and \eqref{eq:Cphi-def} (with the 
basic variables given by \eqref{eq:Ashtekar-B1} instead of \eqref{eq:Ashtekar}), 
where in \eqref{eq:Cphi-def} $p:=|p_1p_2p_3|^{1/3}$.

\subsubsection{Loop quantization}

To quantize the system we follow the program specified in Sect.~\ref{sec:frame-FRW-q},
in particular choosing the polymer representation for the geometry degrees of freedom, while 
keeping the Schroedinger one for the scalar field.

In the geometry part the basic operators are holonomies along straight edges generated
by $\oe^a_i$ and the fluxes across $2$-dimensional rectangles spanned by $\oe^a_i$.
The gravitational kinematical Hilbert space consists of the product of three copies 
of $\Hil_{\gr}$ of the FRW system, each corresponding to one direction of $\oe^a_i$
$\Hil_{\gr} = \bigotimes_{i=1}^3 L^2(\bar{\re},\rd\mu_{\Bohr}) .$
The basis of this space can be built out of eigenstates of the triad (or unit flux) operators
$\hat{p}_i$ and parametrized by three real variables $\lambda_i$ such that
$\hat{p}_i\ket{\lambda_1,\lambda_2,\lambda_3}
=\sgn(\lambda_i)(4\pi\gamma\sqrt{\Delta}\lPl^3)^{\frac{2}{3}}\, \lambda_i^2\,
\ket{\lambda_1,\lambda_2,\lambda_3}$. Alternatively, one of $\lambda_i$ can be 
replaced with the parameter $v:=2\lambda_1\lambda_2\lambda_3$. Here for that 
purpose we select $\lambda_3$, finally labeling the basis elements as 
$\ket{\lambda_1,\lambda_2,v}$.

The space $\Hil_{\phi}$ and the set of basic operators corresponding to the matter 
are the same ones as in the FRW case. The full kinematical Hilbert space is also 
a product $\Hil_{\kin}=\Hil_{\gr}\otimes\Hil_{\phi}$.

The quantum Hamiltonian constraint is constructed out of the classical one by, 
first reexpressing it in terms of the holonomies and fluxes, and next promoting 
these components to operators. In particular the field strength $F^k_{ab}$ is again 
represented via holonomies along closed rectangular loops of the physical area equal to
$\Delta$. Unlike in isotropic case however fixing the loop area does not allow to 
uniquely fix the fiducial lengths of its edges. This apparent ``ambiguity'' gave 
rise to several distinct prescriptions present currently in the literature (including 
the one of \cite{chiou-b1}). On the other hand the relation of the LQC degrees of 
freedom with the full LQG ones constructed in \cite{awe} allowed to fix the relation
uniquely. The constraint resulting from this operation (defined on the dense domain 
in $\Hil_{\kin}\otimes\Hil_{\phi}$) is of the Klein-Gordon form
\begin{equation}\label{eq:constr-b1}
  \hat{C} =\hat{\id} \otimes \partial_{\phi}^2 + \hat{\id}\otimes\Theta_{\B1} .
\end{equation}
Similarly to FRW model we can restrict our interest to just the symmetric sector, that 
is those states $\Psi$, which satisfy $\Psi(\lambda_1,\lambda_2,v,\phi) 
= \Psi(|\lambda_1|,|\lambda_2|,|v|,\phi)$. This allows to restrict the studies 
just to the positive octant $\lambda_1,\lambda_2,v>0$, on which an action of $\Theta_{\B1}$ 
is given by (quite complicated) Eqs.~$(3.35)$-$(3.37)$ of \cite{awe}.
Its important feature is, that, analogously to the isotropic one, it 
divides $\Hil_{\phy}$ onto the superselection sectors built of the states supported 
just on the sets
$\{(\lambda_1,\lambda_2,v); \lambda_1,\lambda_2\in\re, v\in\lat^+_{\varepsilon}\}$ 
with $\lat^+_{\varepsilon} := \{v=\varepsilon+4n;\ n\in\natu\}$, preserved by an 
action of $\Theta_{\B1}$. Therefore to extract the physics one can consider just 
one of those sectors. Here, for simplicity we choose $\varepsilon=0$.

As we show in Appendix~\ref{app:b1-sadj} the operator $\Theta_{\B1}$ admits 
self-adjoint extensions. Knowing its action one can in principle find the physical 
Hilbert space(s) corresponding to the model by analyzing the spectral properties of 
the extensions. On the other hand there exists a well defined procedure of the 
averaging over anisotropies (defined in \cite{awe} to build an embedding of the 
isotropic model in the homogeneous anisotropic one). In this article we will focus 
just on the space of averaged states $\bar{\Hil}_{\phy}$ and their physical properties.

\subsubsection{The isotropic sector}

Following \cite{awe} and the ideas of  \cite{bh-pert} we consider a projection $\pr$ 
mapping from the dense domain in $\Hil_{\gr}$ of the Bianchi I model to the one the isotropic 
model as follows
\begin{equation}
  \psi(\lambda_1,\lambda_2,v) \mapsto \psi(v) = [\pr\psi](v) 
  := \sum_{\lambda_1,\lambda_2}\psi(\lambda_1,\lambda_2,v) .
\end{equation}
Through the direct inspection one can check that there exists an operator 
$\bar{\Theta}_{\B1}$ such that
\begin{equation}
  [\pr \Theta_{\B1}\psi](v) = \bar{\Theta}_{\B1} [\pr \psi](v) .
\end{equation}
An action of that operator equals exactly the one of $\Theta_{\sLQC}$ 
defined via (\ref{eq:ev2-gen},~\ref{eq:fslqc}).

The consequence of the above observation is, that at least to some extent those 
of the aspects of the Bianchi I model, which are related exclusively to the behavior 
of the isotropic degrees of freedom, can be investigated via an isotropic model 
constructed via averaging over anisotropies and equivalent to the sLQC one described 
in Sect.~\ref{sec:frame-FRW}. However, one has to be aware, that some of the physical 
states might in principle be in the kernel of the projection operator $\pr$. Thus, certain
care needs to be taken, when relating the properties of the isotropic sector defined above
with the full Bianchi I model. In particular it is not confirmed, whether the expectation 
values and dispersions of the total volume Bianchi I operators agree with the analogous 
quantities of the volume operators acting on the averaged states. This issue will require 
further studies.

\subsection{Physical Hilbert space, observables} \label{sec:frame-phys}

Known form of the Hamiltonian constraint \eqref{eq:constr-quant} and in particular the 
evolution operator $\Theta$ \eqref{eq:ev2-gen} allows to easily extract the Hilbert 
structure of the space of states annihilated by the constraint. The exact construction
of $\Hil_{\phy}$ is done via group averaging (see \cite{aps-det} for the details). 

To start with, we note, that the spectrum of $\Theta$ is for all the cases considered 
here absolutely continuous \cite{sp-cont}, nondegenerate and equals $\Sp(\Theta)=\re^+\cup \{0\}$
\cite{kl-flat}. 
In consequence one can build a base of $\Hil_{\gr}$ \cite{ind-drop} out of the eigenfunctions 
$e_k$ corresponding to nonnegative eigenvalues
\begin{equation}\label{eq:eigenf}
  [\Theta e_k](v) = \omega^2(k) e_k(v),
\end{equation}
(where $\omega(k) = \sqrt{12\pi G}k$, $k>0$) and normalized such that 
$\ip{e_k}{e_{k'}} = \delta(k'-k)$. In the superselection sectors $\varepsilon=0$ the
remaining freedom of global rotation is furthermore fixed by the requirement, that
\begin{equation}\label{eq:fix-cond}
  e_k(v=4) \in \re^+ .
\end{equation}
Applying the simplest form of the group averaging 
presented in \cite{aps-det} and above spectral decomposition we arrive to the 
following representation of the elements of $\Hil_{\phy}$
\begin{equation}\label{eq:lqc-state}
  \Psi(v,\phi) = \int_{\re^+} \rd k \tilde{\Psi}(k) e_k(v) e^{i\omega(k)\phi} ,
\end{equation}
where $\tilde{\Psi}\in L^2(\re^+,\rd k)$ is the spectral profile of $\Psi$. The 
physical inner product is given by
\begin{equation}\label{eq:phy-ip}
  \ip{\Psi}{\Phi} = \int_{\re^+}\rd k \bar{\tilde{\Psi}}(k)\tilde{\Phi}(k) .
\end{equation}

Since here we are dealing with the constrained system, there is no natural notion of 
time and evolution. It can be provided via the unitary mapping \eqref{eq:evo-map}.
An alternative way to define an evolution is the construction of the family of 
\emph{partial observables} \cite{par-obs}, parametrized by one of the dynamical 
variables and of the elements related via unitary transformation. Here we construct 
family $\ln|\hat{v}|_{\phi}$ \cite{aps-det} interpreted as $\ln|v|$ at given ``time'' 
$\phi$. The systematic way of constructing such observables is presented in \cite{GaSch}. 
For the models considered in the article the expectation values and dispersions of 
$\ln|\hat{v}|_{\phi}$ for the physical state $\Psi$ equal the analogous quantities 
of the kinematical observable $\ln|\hat{v}|$ acting on the initial data
$\psi_{\phi}:=\Psi(\cdot,\phi)\in\Hil_{\kin}$
\begin{equation}\label{eq:lnv}
  [\ln(v)_{\phi_o}\Psi](v) = e^{i\sqrt{\Theta^2}(\phi-\phi_o)} \ln(v)\Psi(v,\phi_o) .
\end{equation}
For completeness we introduce one more observable, corresponding to the constant of 
motion $\ln(\omega)$ -- an operator $\ln(\hat{\omega})$ acting as follows
\begin{equation}\label{eq:lnw}
  [\ln(\hat{\omega}/\sqrt{G})\tilde{\Psi}](k) = \ln(\omega(k)/\sqrt{G})\tilde{\Psi}(k) .
\end{equation}
This operator will be useful later in the paper as (it will be shown that) its dispersion 
bounds the growth of the spread in $\ln|\hat{v}|_{\phi}$.

\subsection{Wheeler-DeWitt analog} \label{sec:frame-WDW}

The systems studied in this article can be also quantized via methods of the geometrodynamics.
Indeed, the geometric component \eqref{eq:Cgr-def} of the Hamiltonian constraint 
can be expressed entirely in terms of the coefficients $(c,p)$ defined in \eqref{eq:Ashtekar}
\begin{equation}\label{eq:Cgr-wdw}
  \ub{C}_{\gr} = -\frac{6}{\gamma^2}c^2\sqrt{p} ,
\end{equation}
and the entire system can be treated just as an abstract one of the phase space
coordinatized by $(c,p,\phi,p_{\phi})$ and quantized via standard methods of quantum 
mechanics. As the result the kinematical Hilbert space takes the form $\ub{\Hil}_{\kin} 
= \ub{\Hil}_{\gr}\otimes L^2(\re,\rd \phi)$, where $\ub{\Hil}_{\gr} := L^2(\re,\rd v)$ 
with the inner product between $\ub{\psi},\ub{\chi}\in\ub{\Hil}_{\gr}$
\begin{equation}\label{eq:gr-ip-wdw}
  \sip{\ub{\psi}}{\ub{\chi}} = \int_{\re} \rd v\, \bar{\ub{\psi}}(v) \ub{\chi}(v) .
\end{equation}
The quantum Hamiltonian constraint can be expressed as a differential analog of 
\eqref{eq:constr-quant}
\begin{equation}\label{eq:C-wdw}\begin{split}
  \partial_{\phi}^2\ub{\Psi}(v,\phi) &= - \ub{\Theta}\,\ub{\Psi}(v,\phi) \\
  &:= 12\pi G\sqrt{|v|}\partial_v|v|\partial_v\sqrt{|v|}\ub{\Psi}(v,\phi) .
\end{split}\end{equation}
To arrive to above equation we selected the factor ordering in \eqref{eq:Cgr-wdw}
consistent with the one of \eqref{eq:Cgr-def} (see \cite{aps-imp} for the detailed 
explanation).

The physical Hilbert space can be again constructed via group averaging and it is
an almost complete analog of the one of LQC models, with just slight differences being 
a consequences of the two-fold degeneracy of the eigenspaces of the operator $\ub{\Theta}$. 
Here the orthonormal basis of (the symmetric states on) $\ub{\Hil}_{\gr}$ consists 
of the functions
\begin{equation}\label{eq:wdw-basis}
  \ub{e}_{k}(v) = \frac{1}{\sqrt{2\pi v}}e^{ik\ln|v|} , \quad k\in\re ,
\end{equation}
and the physical states (positive frequency solutions to \eqref{eq:C-wdw}) have the form
\begin{equation}\label{eq:wdw-state}
  \ub{\Psi}(v,\phi) = \int_{\re} \rd k \tilde{\ub{\Psi}}(k) e_k(v) e^{i\omega(k)\phi} ,
\end{equation}
where $\tilde{\ub{\Psi}}\in L^2(\re,\rd k)$ and $\omega(k)=\sqrt{12\pi G}|k|$. 
The inner product has the same form as \eqref{eq:phy-ip} but now $k$ runs the 
entire $\re$.

To characterize the states and define a physical evolution we use the observables 
$\ln(\hat{\omega})$ and $\ln|\hat{v}|_{\phi}$ given, respectively, by full analogs
of \eqref{eq:lnw} and \eqref{eq:lnv}. The latter ones can be expressed as quite
simple differential operators acting directly on $\ub{\Hil}_{\phy}$ 
\begin{equation}\label{eq:lnv-wdw}\begin{split}
  \ln|\hat{v}|_{\phi}\tilde{\ub{\Psi}} 
  &= -i e^{i\omega(k)\phi} \partial_{k} e^{-i\omega(k)\phi} \tilde{\ub{\Psi}} \\
  &= [-i\partial_{k}-(\partial_k\omega(k))\phi\hat{\id}]\tilde{\ub{\Psi}} .
\end{split}\end{equation}
This fact will be very useful in the following sections, where we will use it to derive 
the relation between the dispersion of the components of the WDW limits of the LQC states.

\section{WDW limit of an LQC state} \label{sec:limit}

The comparison of the forms of the operators $\Theta$ \eqref{eq:ev2-gen} and $\ub{\Theta}$ 
specified via \eqref{eq:C-wdw} shows that under certain conditions (slowly changing 
functions) one of the operators can be approximated by the other. Therefore one may 
expect, that the solutions to \eqref{eq:constr-quant} converge in certain regions to some 
solutions of \eqref{eq:C-wdw}. Indeed, it was shown via numerical methods in \cite{aps-imp}
that the eigenfunctions $e_{k}$ converge to certain combinations of $\ub{e}_k$ and 
$\ub{e}_{-k}$. Furthermore the analytic properties (reality) of $\Theta$ imply that 
this limit has the form of a ``standing wave'' that is it is composed equally of 
incoming $(k>0)$ and outgoing $(k<0)$ plane waves \eqref{eq:wdw-basis}. More precisely 
\cite{remnants}
\begin{equation}\label{eq:WDW-lim-def}\begin{split}
  e_k(v) &= r(k)(e^{i\alpha(k)}\ub{e}_k(v) + e^{-i\alpha(k)}\ub{e}_{-k}(v)) \\
  &\hphantom{=}+ O(|\ub{e}_k(v)|(k/v)^2) \\
  &=: \ub{\psi}_k(v) + O(|\ub{e}_k(v)|(k/v)^2),
\end{split}\end{equation}
where $r(k)\in\re^+$ can be determined analytically via the relations between norms 
of the LQC and WDW states (see Appendix~\ref{app:norms}) and $\alpha(k)\in S^1$ 
is a phase shift. 

In this section we analyze the WDW limits $\ub{\psi}_k$ of the LQC eigenfunctions 
$e_k$ in detail. First, in Sec.~\ref{sec:conv} we provide an analytic proof of the 
convergence for all the forms of $\Theta$ considered in this paper, as well as recall
the arguments allowing to determine the structure of the limit. Second, in 
Sec.~\ref{sec:phases} we perform a detailed analytical and numerical analysis of the 
phase shifts $\alpha(k)$ defined in \eqref{eq:WDW-lim-def}. The properties of these 
shifts are the critical components allowing to arrive to the triangle inequalities
relating the dispersions of the physical state and being the main result of this paper.

\subsection{The convergence of the bases}\label{sec:conv}

In order to explicitly show the convergence \eqref{eq:WDW-lim-def} we compare 
the eigenfunctions $e_k$ -- solutions to \eqref{eq:eigenf}, with the solutions 
\eqref{eq:wdw-basis} to the WDW analog of \eqref{eq:eigenf}. To start with, we note, 
that the Eq.~\eqref{eq:eigenf} is a $2$nd order difference equation, however, due 
to the specific properties of $\Theta$ characteristic for each of the prescriptions 
considered here, the whole solution is determined just by the single value $e_k(v=4)$ 
(see \cite{aps-imp,acs,mmo} for the details on respective prescriptions).

To analyze the solution it is more convenient to rewrite that equation in the $1$st 
order form \cite{simon}. For that we introduce the vector notation, defining
\begin{equation}\label{eq:eig-vec}
  \vec{e}_k(v):= \left( \begin{array}{l} e_k(v) \\ e_k(v-4) \end{array} \right) . 
\end{equation}
Using it we can rewrite \eqref{eq:eigenf} in the following form
\begin{equation}\label{eq:eig-1st}
  \vec{e}_k(v+4) = \blA(v)\,\vec{e}_k(v) ,
\end{equation}
where the matrix $\blA$ can be expressed (with use of the notation introduced in 
Eq.~\eqref{eq:ev2-gen}) as
\begin{equation}\label{eq:A-def}
  \blA(v) = \left( \begin{array}{cc} 
                     \frac{f_o(v)-\omega^2(k)}{f_+(v)} & -\frac{f_-(v)}{f_+(v)} \\
                      1 & 0
                   \end{array} \right) .
\end{equation}
To relate $e_{k}$ with $\ub{e}_{\pm k}$ we note that the value of $e_{k}$ at each
pair of consecutive points $v,v+4$ can be encoded as values of the (specific for the 
chosen pair of points) combination of $\ub{e}_{\pm k}$, that is
\begin{equation}\label{eq:eig-coeff}
  \vec{e}_k(v) = \blB(v) \vec{\chi}_k(v) ,
\end{equation}
where the transformation matrix $\blB$ is defined as follows
\begin{equation}
  \blB(v) := \left( \begin{array}{ll} 
                      \ub{e}_k(v+4) & \ub{e}_k(v+4) \\
                      \ub{e}_k(v) & \ub{e}_k(v)
                    \end{array} \right) .
\end{equation}
Having at hand the objects defined above we can rewrite the equation \eqref{eq:eig-1st}
as the iterative equation for the vectors of coefficients $\vec{\chi}_k$
\begin{equation}\label{eq:M-def}\begin{split}
  \vec{\chi}_k(v+4) &= \blB^{-1}(v) \blA(v) \blB(v-4) \vec{\chi}_k(v) \\
  &=: \blM(v) \vec{\chi}_k(v) .
\end{split}\end{equation}
The exact coefficients of the matrix $\blM(v)$ can be calculated explicitly for 
each of the prescriptions specified in Sect.~\ref{sec:frame-FRW-q}. An important feature
(found by direct inspection) of it is, that in all three cases they have the following 
asymptotic behavior
\begin{equation}
  \blM(v) = \id + \boldsymbol{O}(v^{-3}) , 
\end{equation}
where $\boldsymbol{O}(v^{-n})$ denotes the matrix, all the coefficients of which behave
like $O(v^{-n})$. That convergence implies (via application of the methods presented in 
\cite{simon} Sec.~4) the existence of the limit
\begin{equation}
  \lim_{n\to \infty} \vec{\chi}_k(4n) = \vec{\chi}_k ,
\end{equation}
as well as it confirms the rate of convergence specified in \eqref{eq:WDW-lim-def}. 
This limit can be expressed in terms of the coefficients introduced in \eqref{eq:WDW-lim-def} 
in the following form
\begin{equation}\label{eq:limit-form}
  \vec{\chi}_k = r(k) \left( \begin{array}{l} e^{i\alpha(k)} \\ e^{-i\alpha(k)} \end{array} \right) ,
\end{equation}
which is a consequence of the observation, that all the coefficients $f_{o,\pm}(v)$ 
in \eqref{eq:eigenf} are real, so (by \eqref{eq:fix-cond}) is $e_k$.

The scaling factor $r(k)$ can be easily determined from the relation between
norms in LQC and WDW theory discussed in Appendix~\ref{app:norms} and equals
\begin{equation}\label{eq:rvalue}
  r(k) = 2 .
\end{equation}
The behavior of the phase shift function $\alpha(k)$ is however much less trivial 
and requires detailed studies.

\subsection{The phase shifts}\label{sec:phases}

To extract the properties of the phase shifts $\alpha(k)$ defined in 
\eqref{eq:limit-form} we combine the analytical and numerical methods. 
We focus on the behavior of the derivative $\alpha':=\partial_k\alpha$, as it 
is exactly the quantity which will be relevant in the further studies. First we derive 
analytically the asymptotic behavior of $\alpha(k)$ for $k\to\infty$ for the sLQC
prescription (Sec.~\ref{ssec:as-sLQC}). The analytical results are then strengthened 
and generalized to other prescription by means of the numerical methods described in 
Appendix~\ref{app:num}. The results of that analysis are presented in Sec.~\ref{ssec:as-other}.

\subsubsection{Asymptotics in sLQC}\label{ssec:as-sLQC}

Among the prescriptions considered here the sLQC one is somehow distinguished, as the
bases $e_k$ expressed as functions of an appropriately defined canonical momentum $b$ of $v$ 
have a simple analytical form \cite{acs}. This allows for a quite high level of control 
over their properties, a fact which we will exploit below. To start with, let us fix the 
definition of $b$, choosing it to equal
\begin{equation}
  b := b_o c|p|^{-\frac{1}{2}} , \qquad \{v,b\} = 2 , 
\end{equation}
where $c,p$ are given by \eqref{eq:Ashtekar} and the proportionality constant $b_o$ 
is fixed via the righthand side equality. This quantity can be next used as a configuration 
variable on the quantum level. A particularly convenient choice of the representation 
of the quantum states using this variable is provided by the following transformation operators
chosen respectively for WDW ($\ub{\Fou}$) and LQC ($\Fou$) framework
\begin{subequations}\label{eq:v-b-trans}\begin{align}
  [\ub{\Fou}\psi](b) &= \int_{\re}\rd v\, |v|^{-\frac{1}{2}} e^{\frac{ivb}{2}} \psi(v) , \\
  [\Fou\psi](b) &= \sum_{v\in\lat_{\varepsilon}} |v|^{-\frac{1}{2}} 
  e^{\frac{ivb}{2}} \psi(v) ,
\end{align}\end{subequations}
where the part of $\psi$ supported on $v<0$ is determined by the symmetry requirement.
The form of these transformations implies, that the domain of $b$ is the entire $\re$ in 
the case of WDW and the circle of the radius $1/2$ in LQC.

The evolution operators $\ub{\Theta}$ and $\Theta_{\sLQC}$ transformed via 
\eqref{eq:v-b-trans} take the form
\begin{subequations}\begin{align}
  \ub{\Theta}^2 &= -12\pi G [b\partial_b]^2 , \\
  \Theta^2_{\sLQC} &= -12\pi G [\sin(b)\partial_b]^2 ,
\end{align}\end{subequations}
and their (symmetric) eigenfunctions corresponding to the eigenvalue $\omega^2=12\pi Gk^2$ 
(and in case of sLQC corresponding to the sector $\varepsilon=0$) are combinations
of the orthonormal basis elements
\begin{subequations}\label{eq:b-bases}\begin{align}
  \ub{e}_k(b) &= \ub{N}(k) e^{-ik\ln|b/2|} , \label{eq:b-bases-WdW} \\
  e_k(b) &= N(k) \cos(-k\ln(\tan|b/2|)) , \label{eq:b-bases-sLQC}
\end{align}\end{subequations}
where $\ub{N}(k),N(k)$ are the normalization factors determined by the physical 
inner product \cite{acs} and \eqref{eq:b-bases-sLQC} is written in the chart 
$b\in[-\pi/2,\pi/2]$. The $-$ sign in the exponents comes from the comparison
of the spectrum and bases of the operator $\sqrt{\ub{\Theta}}$ in $v$ and $b$ 
representation (see for example \cite{b1-evo}).

To retrieve the large $v$ behavior of the functions \eqref{eq:b-bases} one needs 
to perform the transform inverse to \eqref{eq:v-b-trans}. Since the large $v$ correspond 
to high frequencies in $b$, the particular form of the functions implies, that only the domain 
near $b=0$ will give the relevant contribution to the asymptotics in $v\to\infty$.

In order to be able to compare the functions $\ub{e}_k(b)$ and $e_k(b)$ one first needs 
to to deal with the fact, that the inverse transform of \eqref{eq:b-bases-sLQC}, 
involves an integration over the domain $[-\pi/2,\pi/2]$ whereas for \eqref{eq:b-bases-WdW}
one need to perform an integration over $\re$.

To do so, let us first consider on $[-\pi/2,\pi/2]$ a function 
  $\xi(b) \ub{e}_k(b),$
where $\xi(b)$ is some smooth function with support in $(-\pi/2,\pi/2)$ and equal 
to $1$ in some neighborhood of $0$. Regarding this function as defined on
the entire $\re$ we note that the difference
\begin{equation}
  g_k(b):=\ub{e}_k(b)-\xi(b) \ub{e}_k(b)
\end{equation}
is a smooth function with appropriate behavior at infinity \cite{smooth-note}. 
Hence its Fourier transform is of the order $O(v^{-N})$ for any $N\in\natu$.

On the other hand, for $\xi(b) \ub{e}_k(b)$ considered as a function on a circle, 
the difference
\begin{equation}
  f_k(b) := e^{-ik\ln(\tan|b/2|)} - \xi(b)e^{-ik\ln|b/2|} .
\end{equation}
is of the class $C_0$, thus by Lebesgue-Riemann lemma its transform $\Fou^{-1}f_k$ 
is of the order $o(|\ub{e}_k(v)|)$. In consequence the function
\begin{equation}
  e'_k(b) = \xi(b)\cos(k\ln|b/2|) + (f_k(b)+f_{-k}(b))/2 ,
\end{equation}
supported on $\re$ has the same WDW limit as $e_k$. Furthermore the components of 
this limit proportional to $\ub{e}_{\pm k}$ correspond to the respective components 
$e^{\mp ik\ln|b/2|}$ of $e'_k$.

Bringing together this two observations we see, that to find the desired phase 
shifts one just needs to find the transform of the functions $\ub{e}_k(b)$. As 
they are the eigenfunctions of $\ub{\Theta}$, they are proportional to $\ub{e}_k(v)$ 
\begin{equation}
  \Fou^{-1}[e^{\mp ik\ln|b/2|}](v) = \tilde{N}(k) \ub{e}_{\pm k}(v)  
\end{equation}
and the factor of proportionality $\tilde{N}(k)$ equals the transform of 
$\sqrt{2\pi}e^{\mp ik\ln|b/2|}$ at the points $v=\pm 1$. Selecting for the component
proportional to $\ub{e}_{k}(b)$ the point $v=-1$ we get
\begin{equation}\label{eq:tN-def}\begin{split}
  \tilde{N}(k) &= {\sqrt{2\pi}}\int_{\re}\rd b\, e^{-i(k \ln|b/2|-b/2)}  \\
  &={\sqrt{8\pi}}\,{k}\,e^{-ik\ln(k)}\int_{\re} \rd y\, e^{-ik(\ln|y| - y)} ,
\end{split}\end{equation}
where to arrive to the latter equality we introduced the change of variables
$b=2ky$. The last integral can be computed in the regime $k\rightarrow \infty$ 
by a stationary phase method (see Appendix~\ref{app:Ftr} for the proof of the 
applicability of the method). The result is
\begin{equation}\begin{split}
  \tilde{N}(k) &\approx {\sqrt{8\pi}}\,{k}\,
  e^{-ik\ln(k)}\sqrt{-2\pi i}\,\left[{\frac{y_o}{\sqrt{k}}} 
  e^{-ik(\ln|y_o|- y_o)}\right]_{y_o=1} \\
  &\approx {4\pi\sqrt{-i}}{\sqrt{k}}\, e^{-i(k\ln(k)-k)}.
\end{split}\end{equation}
Analogously, one can calculate $\tilde{N}(k)$ for the component proportional to 
$\ub{e}_{-k}(b)$ by selecting the point $v=1$. These two results allow us to extract 
the phase shift $\alpha(k)$, which equals
\begin{equation}\label{eq:phase-sLQC}
  \alpha(k) = -k(\ln(k)-1) - \fracs{3}{4}\pi+ o(k^0) .
\end{equation}

Via the same method one can compute the derivative $\partial_k\tilde{N}=:\tilde{N}'$.
\begin{equation}\label{eq:N'int}\begin{split}
  \tilde{N}' &={\sqrt{2\pi}}\int_{\re} \rd b\, [i\ln|b/2|] e^{-i(k\ln|b/2|-b/2)}  \\
  &= {\sqrt{8\pi}}\,{k}\,e^{-ik\ln(k)}\int_{\re}\rd y\, [-i\ln|y|] e^{-ik(\ln|y|-y)}  \\
  &\hphantom{=}- {\sqrt{8\pi}}\,ik\ln(k)\, e^{-ik\ln(k)}\int_{\re} \rd y\, e^{-ik (\ln|y|-y)}  \\
  &\approx  -i\ln(k)\tilde{N}(k) ,
\end{split}\end{equation}
where the last estimate follows from the fact, that decomposing $\tilde{N}(k)=:A(k)e^{i\alpha(k)}$
(where $A(k)\in\re^+$) one can express its derivative as
\begin{equation}\label{eq:alpha-diff}\begin{split}
  \tilde{N}'(k) &= i\alpha'(k) A(k)e^{i\alpha(k)}+A(k)'e^{i\alpha(k)} \\
  &\approx -i{\sqrt{8\pi}\sqrt{-i}}\ln(k)\sqrt{k}\, e^{-i(k\ln(k)-k)} \\
  &\hphantom{=}+ O( k^{-1/2}\ln(k)).
\end{split}\end{equation}
In consequence the phase shift derivative equals
\begin{equation}\label{eq:dphase-sLQC}
  \alpha'(k) = -\ln(k) + O(k^{-1}\ln(k)) .
\end{equation}

\subsubsection{Numerical generalization}\label{ssec:as-other}

In the case of the remaining two prescriptions repeating the analytical calculations 
preformed for the sLQC one is not possible, as the eigenfunctions of $\Theta$ do 
not have manageable analytic form in either of $v$ or $b$ representations. We note 
however, that between the prescriptions the operators $\Theta$ differ just by a compact 
operator. Thus, it is expected that the asymptotic behavior of both $\alpha(k)$ 
and $\alpha'(k)$ corresponding to them is again given by \eqref{eq:phase-sLQC} and 
\eqref{eq:dphase-sLQC} up to the rest terms decaying with $k$. We verify this 
expectation for $\alpha'(k)$, using the numerical methods which are described in detail 
in Appendix~\ref{app:num}. Those methods allow to determine the values of $\alpha'$ 
in quite wide range of $k$ as well as to verify its asymptotic behavior (within 
the limitations of applied numerics). The results for different prescriptions are 
presented in Fig.~\ref{fig:diff1}. Although the exact form of $\alpha'$ depends 
on the prescription, especially for small $k$, one can observe the following 
features common for all of the prescriptions considered in this article:
\begin{enumerate}[(i)]
  \item For large $k$ the derivative $\alpha'$ converges to the limit specified in 
    \eqref{eq:dphase-sLQC} with the rate
    \begin{equation}\label{eq:dalimit}
      \alpha'(k) = -\ln(k) + O(k^{-2}) ,
    \end{equation}
  \item The scaled $2$nd order derivative of $\alpha$ is bounded
    \begin{equation}\label{eq:d2alpha}
      |[k\partial_k^2\alpha](k)| \leq 1
    \end{equation}
    for every value of $k$.
\end{enumerate}
These properties will be crucial for building the relation between the dispersion 
growth through the bounce.

\section{The scattering picture}\label{sec:scatter}

It was shown in Sec.~\ref{sec:limit} that the basis functions spanning the LQC 
physical Hilbert space admit certain WDW limits. Given that one can define a WDW limit of 
any physical state by replacing the basis functions in \eqref{eq:lqc-state} with the 
limits $\ub{\psi}_k$ defined via \eqref{eq:WDW-lim-def}. This operation defines 
a relation between the LQC physical Hilbert space and the WDW one, which in terms of the 
spectral profiles can be written as follows
\begin{equation}\label{eq:phy-lim}
  \tilde{\Psi}(|k|) \mapsto \tilde{\ub{\Psi}}(k) 
  = 2 e^{i\sgn(k)\alpha(|k|)} \sgn(k) \tilde{\Psi}(|k|) ,
\end{equation}
where $k$ spans the entire real line.

That limit consists of two components, the incoming wave packet (corresponding to 
$k>0$) and the outgoing one ($k<0$). On the physical level they represent the 
universe which is, respectively, contracting to big crunch singularity and 
expanding from the big bang one. The entire LQC dynamics can be thus seen as the 
specific kind of ``transition'' between the contracting WDW universe (represented by 
$\ket{\ub{\Psi}}_{\rm in}$) to the expanding 
one (denoted as $\ket{\ub{\Psi}}_{\rm out}$ 
\begin{equation}
  \ket{\ub{\Psi}}_{\rm in} \mapsto \ket{\ub{\Psi}}_{\rm out} 
  = \hat{\rho} \ket{\ub{\Psi}}_{\rm in} ,
\end{equation}
In consequence, looking at the evolution ``from the infinity'' (in the configuration 
space or in cosmic time) one can interpret the evolution as the process of scattering
of the contracting geometrodynamical universe. The form of the limit \eqref{eq:phy-lim} 
immediately allows to write down the scattering matrix
\begin{equation}
  \rho(k,k') = (\ub{e}_k|\hat{\rho}|\ub{e}_{k'}) 
  = e^{-\sgn(k')\alpha(|k'|)}\delta(k+k') ,
\end{equation}
which form encodes in particular the fact, that the contracting universe totally 
``reflects'' into the expanding one
\begin{equation}\label{eq:scatt}
  \tilde{\ub{\Psi}}(k) \mapsto e^{2\sgn(k')\alpha(|k'|)} \tilde{\ub{\Psi}}(-k) .
\end{equation}
This picture allows to address in a quite natural and intuitive way the questions 
regarding the relation of the properties of the pre bounce branch (universe in the 
asymptotic past) and the post bounce one (asymptotic future). In particular, we 
will apply it in the next section to determine how much the bouncing universe 
can disperse in the distant future of the bounce in comparison to the initial spread 
in its distant past.

When considering the above picture one has to remember that, although the LQC basis
functions converge to the combinations of the WDW ones, this is not necessarily the 
truth for the general physical states, as the convergence of the bases is not uniform
with respect to $k$. Nonetheless, once the attention is restricted just to the states
localized with respect to the observable $\hat{k}$ defined analogously to \eqref{eq:lnw}
(that is of the finite dispersion in $k$) the uniformity is restored and the WDW limit
is defined in the precise sense. This fact is used for example in Sec.~\ref{sec:triangle-wdw}
where the expectation values and dispersions of the LQC states are related with the ones
of its WDW limit. 

The scattering picture can obviously be constructed in the context of any LQC model 
in which the basis functions converge explicitly to the WDW ones, like for example the
models with the positive cosmological constant \cite{posL} or the Bianchi I ones 
\cite{b1-evo}. The applicability is however not restricted just to such systems. 
In particular the models featuring the classical recollapse, like \cite{spher,negL}, in LQC
admit a quasi-periodic evolution. For those models it is also possible to build 
a correspondence between the LQC and WDW states, since the basis elements of $\Hil_{\phy}$
converge to their WDW analogs also there. The new difficulty in these cases is the fact
that, as the spectra of the LQC evolution operators are discrete while the WDW ones 
are continuous, the direct analog of the transformation \eqref{eq:phy-lim} leads 
to the WDW states of the zero norm. This problem can be circumvented by introducing 
an appropriate interpolation of the discrete spectral profile of an LQC state. 
The WDW state constructed this way represents a single epoch (between the bounces) 
of the evolution of the LQC one. However, since such interpolations are not defined 
uniquely, there is no direct $1-1$ correspondence between the loop states and their 
``limits''. One can however choose the interpolations which reproduce the expectation 
values and the dispersions of the relevant observables at least approximately. This 
way it is possible build the WDW states well mimicking one epoch of the loop universe 
evolution. It happens however at the cost of relaxing the relations between the 
physical parameters corresponding to them to approximate ones, without explicit 
convergence of their values. The reason for that is two-fold:
\begin{enumerate}[(i)]
  \item nontrivial corrections due to interpolation of the discrete spectral profile, and
  \item the fact that the basis elements of the LQC physical Hilbert space converge
    to their (rescaled) WDW analogs only asymptotically, thus obviously beyond the
    classical recollapse point.
\end{enumerate}
Despite this, such relations can be still quite useful. In particular this method is
well suited to address the question, how the parameters (for example dispersions)
can change between the epochs. In particular it can be used to investigate the 
issue of the spontaneous coherence of the LQC state, that is to address the question 
whether, given an initial date describing the state which is not semiclassical, 
the state will admit in the future evolution the semiclassical epoch.

\section{The dispersion analysis}\label{sec:triangle}

This section is dedicated to the main goal of this paper: finding the precise relation
between the dispersions of the physical LQC state representing the universe in the 
distant future (post bounce) and past (pre bounce). The studies are divided onto 
two steps. First, in Sec.~\ref{sec:triangle-wdw} we apply the scattering picture 
to relate the dispersions of incoming and outgoing asymptotic WDW states. Found relation
is next translated in Sec.~\ref{sec:tr-rel} to obtain the relation between the dispersions 
of a genuine LQC state in the asymptotic past and future.

\subsection{Dispersions of the WDW limit components}\label{sec:triangle-wdw}

Given the WDW limit $\ub{\Psi}$ (defined by \eqref{eq:phy-lim}) of the LQC state described 
by the spectral profile $\tilde{\Psi}$ let us define its decomposition onto the incoming 
$\ub{\Psi}^+$ and outgoing $\ub{\Psi}^-$ components such that the
spectral profiles corresponding to them equal
\begin{equation}
  \tilde{\ub{\Psi}}^{\pm}(k) := \theta(\pm k) \tilde{\ub{\Psi}}(k) , 
\end{equation}
where $\theta$ is a Heaviside step function. Denote the subspaces of $\ub{\Hil}_{\phy}$ 
formed by these components as $\ub{\Hil}^{\pm}_{\phy}$ respectively. On each of these 
components one can consider an action of the observables $\ln|v|_{\phi}$ defined by 
\eqref{eq:lnv-wdw}. 
Their expectation values and dispersions equal respectively
\begin{subequations}\label{eq:trajs}\begin{align}
  v(\phi) &:= \langle \ln|v|_{\phi} \rangle_{\pm} 
  = \pm a\,\phi\|\ub{\Psi}^{\pm}\| + \langle -i\partial_k \rangle_{\pm} , \\
  \sigma_{\pm} &:= \langle\Delta \ln|v|_{\phi} \rangle_{\pm}
  = \langle\Delta (-i\partial_k) \rangle_{\pm} , \label{eq:disp-vals}
\end{align}\end{subequations}
where $a:=\sqrt{12\pi G}$ and for any observable $\hat{O}$ we define 
$\langle \hat{O} \rangle_{\pm} := \bra{\ub{\Psi}^\pm} \hat{O} \ket{\ub{\Psi}^\pm}$.

The main question we would like to address here is whether there exists
the relation between $\sigma_-$ and $\sigma_+$ and what is its form. 
The answer to the former is certainly true as the transformation \eqref{eq:scatt}
unitarily maps $\ub{\Psi}^+ \to \ub{\Psi}^-$ in the following way
\begin{equation}\label{eq:umap}\begin{split}
  &\tilde{\ub{\Psi}}^-(k) = [U\tilde{\ub{\Psi}}^+](k) = e^{-2\alpha(|k|)}\tilde{\ub{\Psi}}^+(-k) , \\
  &U:\ub{\Hil}^+_{\phy} \to \ub{\Hil}^-_{\phy} ,
\end{split}\end{equation}
thus the expectation values and dispersions in \eqref{eq:trajs} are related as 
follows
\begin{subequations}\label{eq:exp-trans}\begin{align}
  \langle -i\partial_k \rangle_{-} &= \langle U^{-1}[-i\partial_k]U \rangle_{+} , \\
  \langle\Delta [-i\partial_k] \rangle_{-} 
  &= \langle\Delta U^{-1}[-i\partial_k]U \rangle_{+} , \label{eq:disp-trans}
\end{align}\end{subequations}
where
\begin{equation}\label{eq:op-split}
  U^{-1}[-i\partial_k]U = -i\partial_k - 2\alpha'\id . 
\end{equation}
Combining together \eqref{eq:disp-vals}, \eqref{eq:umap}, \eqref{eq:disp-trans}, 
\eqref{eq:op-split} and applying very general bound on the dispersion of the sum 
of operators \eqref{eq:disp-sum} we obtain the following inequality
\begin{equation}
  \sigma_- \leq \sigma_+ + 2\langle\Delta \alpha'\id\rangle_+ . 
\end{equation}
To write it down in the useful form we have to express the quantity 
$2\langle\Delta \alpha'\id\rangle_+$ in terms of dispersions of observables commonly 
used to characterize the physical properties of the state. For that we exploit the 
properties of the function $\alpha'$ found in Sect.~\ref{sec:phases}. Namely, by 
the definition of the dispersion we can bound the term under consideration via
\begin{equation}\label{eq:interm1}
  \langle\Delta \alpha'\id\rangle_+^2 
  = \langle ({\alpha'}^2 - \langle\alpha'\rangle_+)^2\,\id\rangle_+
  \leq \langle ({\alpha'}^2 - {\alpha'}^\star)^2\,\id\rangle_+ ,
\end{equation}
which is true for any value of ${\alpha'}^\star$. Here we choose it to be
\begin{equation}
  {\alpha'}^{\star} = \alpha'(\exp(\lambda^\star)) ,\quad
  \lambda^{\star} := \langle\ln(k)\rangle_+ .
\end{equation}
Upon that choice, applying the property \eqref{eq:d2alpha} of $\alpha$ we can 
bound the left-hand side of \eqref{eq:interm1} as follows
\begin{equation}\label{eq:interm2}
  \langle\Delta \alpha'\id\rangle_+^2 
  \leq \langle (\ln(\hat{k}) - \lambda^\star \id)^2 \rangle_+ 
  = \langle \Delta \ln(\hat{k}) \rangle_+^2 .
\end{equation}
Finally, knowing the relation $\omega(k)$ we can express the right-hand side of
\eqref{eq:interm2} via the dispersion of the WDW analog of the observable \eqref{eq:lnw},
which corresponds just to a logarithmic scalar field momentum $\ln(p_{\phi}/b)$, 
where $b:=\hbar\sqrt{G}$. The result is
\begin{equation}\label{eq:triangle1}
  \sigma_- \leq \sigma_+ + 2\sigma_{\star},
\end{equation}
where 
\begin{equation}
  \sigma_{\star} = \langle\Delta\ln(\hat{p}_{\phi})/b\rangle_+
  = \bra{\Psi} \Delta \ln(\hat{p}_{\phi})/b \ket{\Psi} .
\end{equation}
The righthand equality is here a direct consequence of the form of operator 
\eqref{eq:lnw} (multiplication operator in $k$) and the transformation \eqref{eq:phy-lim}.

One can easily see, that the role of $\sigma_-$ and $\sigma_+$ can be exchanged. 
The only modification induced by this operation will be the exchange 
$U\leftrightarrow U^{-1}$. In consequence, \eqref{eq:triangle1} is supplemented
by the inequality
\begin{equation}\label{eq:triangle2}
  \sigma_+ \leq \sigma_- + 2\sigma_{\star},
\end{equation}
thus both these inequalities \eqref{eq:triangle1} and \eqref{eq:triangle1} can 
be understood as the triangle inequalities.

Note that to arrive to above inequalities we have not assumed any semiclassicality 
conditions in any epoch of the state evolution, neither we required the state to be peaked 
about any trajectory. The relations hold for every element of $\Hil_{\phy}$. 

It is also worth noting, that although \eqref{eq:triangle1}, \eqref{eq:triangle1}
are formulated in terms of the dispersions of the logarithmic observables 
$\ln|\hat{v}|_{\phi}$ and $\ln(\hat{p}_{\phi}/b)$, for the states semiclassical 
(sharply peaked) in any epoch of the evolution (pre or post bounce) 
these quantities can be approximated via analogous ``linear'' ones: 
$|\hat{v}|_{\phi}$, $\hat{p}_{\phi}$
\begin{subequations}\label{eq:lin-obs}\begin{align}
  \langle \ln|\hat{v}|_{\phi} \rangle &\approx \ln \langle |\hat{v}|_{\phi} \rangle , &
  \langle \ln(\hat{p}_{\phi}/b) \rangle 
    &\approx \ln(\langle\hat{p}_{\phi}\rangle/b) , \notag \\
  \langle\Delta \ln|\hat{v}|_{\phi} \rangle 
    &\approx \frac{\langle\Delta |\hat{v}|_{\phi} \rangle}{\langle |\hat{v}|_{\phi} \rangle} ,  &
  \langle \ln(\hat{p}_{\phi}/b) \rangle 
    &\approx \frac{\langle\Delta\hat{p}_{\phi}\rangle}{\langle\hat{p}_{\phi}\rangle} .
  \tag{\ref{eq:lin-obs}}
\end{align}\end{subequations}
In consequence the inequalities \eqref{eq:triangle1}, \eqref{eq:triangle2} can 
be reformulated in terms of them at least on the semi-heuristic level (or in precise 
sense under certain additional assumptions imposed on the form of the state). This 
is not however the aim of the article, as we intended to find a relation which is 
maximally general while remaining precise.

At this point one has to be aware of an important subtlety related to the description 
method applied hare. Namely, the considered observables are the geometrodynamical 
observables acting on the asymptotic states (wave packets), not the exact LQC observables
acting on the LQC states. Therefore one may in principle worry, that due to some wild 
behavior of the LQC basis functions near the bounce point there might be some residual
contributions to the results (expectation values, dispersions) of the scattering picture
essentially invalidating found inequalities, once applied to exact LQC observables.

On the other hand, the studies of \cite{acs} performed for ``linear'' observables 
$|v|_{\phi}$ show, that at least for the sLQC prescription the LQC dispersions indeed 
approach the ones of WDW limits. This suggests that the problematic corrections mentioned
above are not sufficient to distort the main results. Indeed, one can confirm this 
expectations in quite general setting using the relation of the norms of LQC state 
and its WDW limit derived in Appendix~\ref{app:norms}.

\subsection{The relation with LQC observables}\label{sec:tr-rel}

To show it let us focus our attention to just one limit, say in the distant past. Due to 
the symmetry of the system the reasoning is immediately applicable also to the distant 
future one. Also, since the relation is of physical interest only when $\sigma_{\pm}$ stay finite
we restrict our studies to the states $\Psi$ which are localized in the weak sense, 
that is for which the expectation values and dispersions of the components $\ub{\Psi}^{\pm}$
of their WDW limit are finite (for finite $\phi$).

The forms \eqref{eq:lqc-state}, \eqref{eq:wdw-state} of the LQC and WDW states and 
the relation between $\Psi$ and its limit given by \eqref{eq:limit-form} and \eqref{eq:rvalue} 
imply immediately that
\begin{equation}\label{eq:norm-eq}
  \| \ub{\Psi}^+ \| = 2\| \Psi \| .
\end{equation}
On the other hand, via the mapping \eqref{eq:evo-map} these physical norms can be 
expressed as the appropriate kinematical norms on the surface $\phi=\phi_o=\const$, 
which are given by \eqref{eq:gr-ip} and \eqref{eq:gr-ip-wdw} respectively. This 
allows us to define the \emph{partial norms} $\| \cdot \|^{\pm}_{(x,\phi_o)}$ as the 
restrictions of \eqref{eq:gr-ip}, \eqref{eq:gr-ip-wdw} to those points in the domains 
which satisfy $\ln|v|>x$ for '+' and $\ln|v|<x$ for '-' respectively.

Consider now an arbitrary small $\epsilon>0$ and select the point $x_o$ such that
$\|\ub{\Psi}^+\|^-_{(x_o,\phi_o)} \leq \epsilon \|\ub{\Psi}^+\|$. Next define
the function 
\begin{equation}\label{eq:tx-def}
  \tilde{x}(\phi) = x_o - (\alpha/2)(\phi-\phi_o).
\end{equation}
Using this function as a separator we can define the \emph{partial dispersions} 
$\sigma^{\pm}_{\phi}$ of the observables $\ln|\hat{v}|_{\phi}$ per analogy to the 
partial norms, that is restricting the domains of summation to the sets 
$\ln|v| > \tilde{x}(\phi)$, $\ln|v| < \tilde{x}(\phi)$ respectively. In the similar 
way we define the dispersions $\ub{\sigma}^{\pm}_{\phi}$ of analogous observable 
acting on the state $\ub{\Psi}^+$. They obviously sum up to the complete norms and 
dispersions
\begin{subequations}\begin{align}
  \| \Psi \|^2 &= (\| \Psi \|^+_{\phi})^2 + (\| \Psi \|^-_{\phi})^2, \label{eq:lqcn-sum}\\
  \sigma^2_{\phi} &= (\sigma^{+}_{\phi})^2 + (\sigma^{-}_{\phi})^2 , \label{eq:lqcs-sum}\\
  \| \ub{\Psi}^+ \|^2 &= (\| \ub{\Psi}^+ \|^+_{\phi})^2 + (\| \ub{\Psi}^+ \|^-_{\phi})^2 , \label{eq:wdwn-sum}\\
  \sigma^2_+ &= (\ub{\sigma}^{+}_{\phi})^2 + (\ub{\sigma}^{-}_{\phi})^2, \label{eq:wdws-sum}
\end{align}\end{subequations}
where $\sigma_{\phi}$ is the dispersion of an observable \eqref{eq:lnv} and $\sigma_+$ 
is defined via \eqref{eq:disp-vals}.
Using the known asymptotic behavior of the basis functions we can compare the limits 
of these dispersions as $\phi\to -\infty$.

Let us start with $\sigma^+_{\phi}$. Applying the numerical estimate \eqref{eq:WDW-lim-def}
and taking into account, that the state $\Psi$ is localized in $\hat{p}_{\phi}$ we can write
the asymptotic behavior of the wave function (for large negative $\phi$)
\begin{equation}\label{eq:psi-conv}
  \Psi(v,\phi) = \ub{\Psi}^+(v,\phi) + O_1(v^{-5/2}) ,
\end{equation}
where the remnant $O_1$ has the bound independent of $\phi$. This, together with 
the fact that the value $\ub{\Psi}^+(v,\phi)$ depends only on $\ln|v|-\alpha\phi$ 
and the convergence of the sum over the relevant part of $\lat^+_{0}$ and integral
over the domain $\ln|v|>\tilde{x}(\phi)$ implies
\begin{subequations}\label{eq:ndp-conv}\begin{align}
  \|\Psi\|^{+}_{\phi} &= \fracs{1}{2}\|\ub{\Psi}^{+}\|^{+}_{\phi} + O_2(e^{-\tilde{x}(\phi)}) , 
    \label{eq:normp-conv}\\
  \sigma^+_{\phi} &= \ub{\sigma}^+_{\phi} + O_3(\tilde{x}(\phi)e^{-\tilde{x}(\phi)}) .
    \label{eq:sp-conv}
\end{align}\end{subequations}

On the other hand, the localization of $\ub{\Psi}^+$ in $\ln|v|_{\phi}$ and the fact, that
\begin{equation}\label{eq:traj-diff}
  | \bar{x}(\phi) - \tilde{x}(\phi) | \propto \tilde{x}(\phi), \quad
  \bar{x}(\phi) := \frac{\bra{\ub{\Psi}^+} \ln|v|_{\phi} \ket{\ub{\Psi}^+}}{\|\ub{\Psi}^+\|^2},
\end{equation}
implies that
\begin{equation}
  \lim_{\phi\to -\infty} \ub{\sigma}^-_{\phi} = 0, \quad
  \lim_{\phi\to -\infty} \| \ub{\Psi}^+ \|^-_{\phi} \tilde{x}(\phi) = 0 ,
\end{equation} 
where the right-hand side equality is the consequence of the left-hand side one and
\eqref{eq:traj-diff}. Furthermore, from that and \eqref{eq:wdws-sum} follows
\begin{equation}\label{eq:wdw-sp-lim}
  \lim_{\phi\to -\infty} \ub{\sigma}^{+}_{\phi} = \sigma_+ .
\end{equation}
Combining together \eqref{eq:norm-eq}, \eqref{eq:wdwn-sum}, \eqref{eq:sp-conv} and 
\eqref{eq:lqcn-sum}, taking into account, that the part of $\Psi$ contributing 
to the norms and dispersions is supported on $\ln|v|>0$, estimating the partial 
dispersion by the partial norm 
\begin{equation}
  \sigma^-_{\phi} \leq \|\Psi\|^-_{\phi}\, 
  \sup_{v\in\lat^0 \cap [0,\exp(\tilde{x}(\phi))] }  |\ln|v|-\bar{x}(\phi)|
\end{equation}
and applying \eqref{eq:traj-diff} we get
\begin{equation}
  \lim_{\phi\to -\infty} \sigma^{-}_{\phi} = 0.
\end{equation}
This result, together with \eqref{eq:lqcs-sum}, \eqref{eq:sp-conv}, \eqref{eq:tx-def} 
and \eqref{eq:wdw-sp-lim} finally implies
\begin{equation}
  \lim_{\phi\to -\infty} \sigma_{\phi} = \sigma_{+} .
\end{equation}
Due to symmetry of the system, repeating the above reasoning for the limit $\phi\to\infty$
and the WDW component $\ub{\Psi}^-$ we obtain the convergence in the asymptotic future
\begin{equation}
  \lim_{\phi\to +\infty} \sigma_{\phi} = \sigma_{-} ,
\end{equation}
where $\sigma_{-}$ is defined via \eqref{eq:disp-vals}. Thus, provided the considered 
LQC state $\Psi$ has finite dispersion in $\hat{p}_{\phi}$, the triangle inequalities 
\eqref{eq:triangle1}, \eqref{eq:triangle2} apply also to the asymptotic past and 
future limits of the dispersions of LQC states.

\section{Conclusions and outlook}\label{sec:concl}

In this article we introduced the interpretation of an evolution of a universe described 
via Loop Quantum Cosmology as the scattering process of the geometrodynamical (Wheeler 
DeWitt) one. Using this picture we analyzed certain properties of the bounce in 
the model of a flat isotropic universe with the massless scalar field as well as 
in the isotropic sector of its generalization to the homogeneous but anisotropic 
spacetimes (Bianchi I). In these cases the LQC evolution is a process of transition
of an ever-contracting (incoming) WDW universe into an ever-expanding (outgoing) one
per analogy with the Klein-Gordon wave packet coming from the infinity, interacting 
with the nontrivial potential and being reflected in a scattering process back to infinity.
The process is described by a scattering matrix, which explicit form was found and 
of which properties were investigated in detail. These properties were used to compare
the dispersions of the observables $\ln|v|_{\phi}$ -- scaled logarithmic volume at given
``moment'' of a scalar field (an internal time) of the incoming and outgoing states.
It was proved that the dispersions satisfy certain triangle inequalities (\ref{eq:triangle1}, 
\ref{eq:triangle2}) involving also the spread of the logarithmic value of a scalar 
field momentum $\ln(p_{\phi}/(\hbar\sqrt{G}))$. The derived inequalities are: $(i)$ 
exact and, $(ii)$ general, as they hold for \emph{any physical state} admitted by 
the model.

These results were immediately extended to the infinite past and future limits of 
the dispersions of true observables acting on the genuine LQC states, as it was shown, 
that these dispersions converge in the asymptotic past and future to the appropriate 
dispersions of the, respectively, incoming and outgoing WDW states considered in 
the scattering picture. This convergence happens for every state on which the 
dispersion $\sigma_{p_{\phi}}$ of $\hat{p}_{\phi}$ is finite.

The result reported above immediately implies that once the incoming state is semiclassical
(that is it is sharply peaked in $\ln|v|_{\phi}$ and $\ln(p_{\phi}/(\hbar\sqrt{G}))$) so is
the outgoing one and vice versa. In consequence \emph{any physical} universe of finite
$\sigma_{p_{\phi}}$ semiclassical in the infinite past is also \emph{always} semiclassical 
in the infinite future. This conclusion is a precise confirmation as well a generalization 
of the results of \cite{cs} as it $(i)$ holds for any physical state satisfying just a very 
reasonable condition of the finiteness of $\sigma_{p_{\phi}}$, $(ii)$ is valid also for the 
prescriptions different that sLQC and which in particular are not exactly solvable, 
$(iii)$ extends immediately also to the isotropic sector of the Bianchi I model quantized in 
\cite{awe}.

In addition to confirming the robustness of the recall picture for the simplest 
models, the presented work constitutes the development of a new methodology, which 
application is not restricted to exactly solvable models. Our method uses only the 
structure of the physical Hilbert space and asymptotic properties of its basis, 
none of which have to be controlled analytically. The only relevant requirement 
for application is sufficiently good control over states in the geometrodynamical 
analog of the LQC model under study. This allows in particular to easy extend our 
studies to more complicated isotropic models admitting well defined WDW limit, for 
example the ones with positive cosmological constant. For such models by construction
our method seems to be better suitable to investigate asymptotic behavior (very large 
size) of the universe than the ones using the Hamburger decomposition \cite{boj-ham}, 
as in this regime they reach the limit of their applicability \cite{GMpriv}.

Presented methodology can be extended also to any system, in which one, while not 
having a good control over the large $v$ limit of the LQC state, one still can 
verify the semiclassicality preservation of its WDW analog. There, instead of 
comparing the incoming and outgoing components $\ub{\Psi}^{\pm}$ of the WDW limit
at $v\to\infty$, one can compare them in the limit $v\to 0$, where they usually 
converge to the plane waves characteristic for the model we studied in this article. 
Such comparison alone is not sufficient to give any useful information about actual 
dispersions in context of LQC, however it provides a controllable bridge between 
the WDW components. 
Then, once we are able to control (bound) the growth of dispersions along the evolution
of the WDW states, we can use this bridge to relate the dispersions at large 
$v$ of $\ub{\Psi}^+$ and $\ub{\Psi}^-$. Such method seems to be viable for example 
in the case of the universe with massive scalar field (i.e. the inflaton 
\cite{infl}). There, the WDW model is much easier to handle as, in opposition to the 
LQC one, it admits a good internal clock, evolution along which is generated by 
a self-adjoint operator.

In certain, less precise sense an extension can be made also in the case of the 
recollapsing models. There however, as in the physically interesting domain (universe 
sizes not bigger than the size at the recollapse) the WDW limit of the state only 
approximately approaches the genuine LQC one and the real convergence happens already 
for the ``tails'' of the wave function, the parameters (like dispersions) of the 
WDW limits can resemble the analogous ones of LQC state only approximately, without 
actual convergence. Therefore any found relation between the dispersions of the 
components of the limit can provide analogous relation for the LQC state only at 
the approximate level, up to some finite corrections. To get exact relations, 
like the triangle inequalities derived here, the precise estimates on those corrections 
need to be made. On the other hand, the presented scattering interpretation is well fit 
to investigate in systems featuring quasi-periodic dynamics the phenomena like 
spontaneous coherence in some epochs of the universe evolution.

The methodology described in this article can be applied not only to the isotropic 
models, but also to the less symmetric ones. Examples of such are Bianchi I models: 
vacuum ones or admitting the scalar field. In particular the analysis of the latter 
model, quantized via the methods of \cite{chiou-b1} and \cite{szulc-b1} will be 
presented in \cite{B1disp}. There, the semiclassicality preservation has a slightly 
weaker sense, as, due to nontrivial dispersion relation $\omega(\vec{k})$ only the 
relative logarithmic observables $\langle\Delta\ln|v_i|_{\phi}\rangle/\langle\ln|v_i|_{\phi}\rangle$
have well defined finite limits as $\phi\to\pm\infty$. The results describing the semiclassicality
preservation in the vacuum case (quantized as in \cite{b1-mmp}) are already presented 
in \cite{b1-evo}. There, due to subtleties related with the choice of an emergent time the
analogs of triangle inequalities contain additional terms, thus the relations are weaker than
in the case with the scalar field.

\begin{acknowledgments}
  We would like to thank Fernando Barbero, Jerzy Lewandowski and Guillermo Mena Marug\'an 
  for helpful comments.
  The work was supported in part by the Spanish MICINN project no FIS2008-06078-C0303,
  the Polish Ministerstwo Nauki i Szkolnictwa Wy\.zszego grants No.~1~P03B~075~29, 
  No.~182/N-QGG/2008/0 and the Foundation for Polish Science grant Master. 
  TP acknowledges also the financial aid by the I3P program of CSIC and the European 
  Social Fund, and the funds of the European Research Council under short visit 
  grants No.~2127 and No.~3024 of QG network.
\end{acknowledgments}

\appendix

\section{Mathematical aspects of the analysis}\label{app:math}

In this Appendix we discuss certain aspects of the analysis, which, while are applied 
in our studies, are not directly related to the topic of investigation. We also recall 
some general features of quantum mechanics which were used in the main body of the paper.

\subsection{The existence of self-adjoint extensions of Bianchi I evolution operator}
\label{app:b1-sadj}

The evolution operator $\Theta_{\B1}$ appearing in \eqref{eq:constr-b1} is a symmetric 
operator in the appropriate domain dense in $\Hil_{\kin}$. In order to define the 
physical Hilbert space it has however to admit at least one self-adjoint extension. 
Here, applying the analysis of the deficiency spaces \cite{SiRe} we show, that this 
is indeed the case.

The deficiency spaces $\mathcal{U}^{\pm}$ are the spaces of kinematically normalizable 
solutions $\psi^{\pm}$ to the equation
\begin{equation}\label{eq:defic}
  \Theta_{\B1}\psi^{\pm} = \pm i \psi^{\pm} .
\end{equation}
The existence of the self-adjoint extensions depends on the dimensionality of 
$\mathcal{U}^{\pm}$: if $\dim(U^+)=\dim(U^-)$ the operator admits the desired 
extensions. In particular if both the spaces are trivial, the extension is unique.

At present the complicated structure of $\Theta_{\B1}$ makes finding the solutions 
to \eqref{eq:defic} very difficult, however to achieve the task at hand one just 
needs to demonstrate the equality of $\dim{\mathcal{U}^\pm}$. To show that we 
construct a $1-1$ correspondence between the elements of these two spaces.

Suppose, that $\psi^+$ is the element of $\mathcal{U}^+$, that is it satisfies the equation
$\bra{\lambda_1,\lambda_2,v}\Theta_{\B1}-i\id\ket{\psi^+} = 0$ for every basis element
$\bra{\lambda_1,\lambda_2,v}$. Expanding this set of equations in terms of 
$\psi^+(\lambda_1,\lambda_2,v)$, acting on it with complex conjugate and recomposing again
one can see immediately (by inspection of the Eqs.~$(3.35)$-$(3.37)$ of \cite{awe}), 
that $\bar{\psi}^+$ is the solution to $\bra{\lambda_1,\lambda_2,v}\Theta'^2+i\id\ket{\bar{\psi}^+} = 0$, 
that is it belongs to $\mathcal{U}^-$. Since this reasoning can be also repeated in 
the opposite direction, the transformation 
\begin{equation}
  R:\mathcal{U}^+\to\mathcal{U}^-:\ 
  [R\psi^+](\lambda_1,\lambda_2,v) = \bar{\psi}^+(\lambda_1,\lambda_2,v) 
\end{equation}
is a bijection. In consequence the dimensions of $\mathcal{U}^\pm$ are indeed equal.

\subsection{Relation of norms of the LQC states and their WDW limits}
\label{app:norms}

In most LQC quantization prescriptions considered in the literature the basis functions 
$e_k$ -- normalized eigenfunctions of $\Theta$, can be evaluated only numerically. 
This is done by solving the difference equation \eqref{eq:eigenf} for some chosen 
initial data $e_k(v=\varepsilon)$. One does not know however, for which value of
$e_k(\varepsilon)$ the solution is precisely normalized. Therefore in numerical 
studies one usually calculates the eigenfunctions which are not normalized, evaluates
their norm and rescales them appropriately. However in many models, like the one 
considered here, the basis functions are normalized to Dirac $\delta$, thus the 
norm cannot be computed by purely numerical methods. Fortunately, the self-adjointness 
of $\Theta$ implies quite simple relation between the norm of any eigenfunction of 
$\Theta$ and the norm of its WDW limit. Since for given eigenfunction this limit 
can be calculated numerically (see \cite{aps-imp} and Appendix~\ref{app:num}), that 
relation provides sufficient data to normalize the LQC basis functions. Such method 
was implemented for example in \cite{aps-det,aps-imp} and \cite{b1-mmp}. Although 
the discussed relation was applied already in those papers, due to lack of space 
its derivation was never presented. We show it here, since that relation is a key 
ingredient applied in the studies of Sec.~\ref{sec:tr-rel}.

The derivation is essentially an estimate of the product $\ip{e_{k'}}{e_k}$ by the
products $\ip{\ub{e}_{\pm k'}}{\ub{e}_{\pm k}}$ via use of the asymptotic relation
\eqref{eq:WDW-lim-def}. For simplicity we restrict the derivation just to the case 
$\varepsilon=0$, restricting the support of the eigenfunction to $\lat^+_0$, although 
it immediately extends to the remaining superselection sectors. The only difference 
in these sectors is the need to take into account both the limits $v\to\pm\infty$ 
in some prescriptions. The intermediate relations presented above are well defined 
in the distributional sense.

Let us start with the orthonormality condition for the LQC basis functions
\begin{equation}
  X(k,k') := \ip{e_{k'}}{e_k} = \sum_{\lat^+_0} \overline{e_{k'}(v)}e_k(v) = \delta(k-k') .
\end{equation}
She sum in the above equality can be spit as follows
\begin{equation}
  X(k,k') = F_1(k,k') + \sum_{\lat_1} \overline{e_{k'}(v)}e_k(v)
\end{equation}
where $\lat_1 := \lat^+_0|_{v\geq 1}$ and
\begin{equation}
  F_1(k,k') := \sum_{\lat^+_0\cap [0,1]} \overline{e_{k'}(v)}e_k(v)
\end{equation}
vanishes in the sector under consideration and is well defined function in the 
general situation. Applying the limit \eqref{eq:WDW-lim-def} and taking into account, 
that the terms containing the remnant parts will always sum up to a finite quantity 
we get
\begin{equation}\begin{split}
  &X(k,k') = F_2(k,k') + r(k)r(k') \times \\
  &\times \sum_{s,s'=\pm 1} e^{i(s\alpha(k)-s'\alpha(k'))}  
  \sum_{\lat_1} \overline{\ub{e}_{s'k'}(v)}\ub{e}_{sk}(v)
\end{split}\end{equation}
The sum over $\lat_1$ can be now estimated via integral $\int_1^\infty\rd v$ \cite{norm-symm}. 
The form \eqref{eq:wdw-basis} of $\ub{e}_{\pm k}$ implies that the correction due 
to this estimate is again well defined function of $k,k'$, thus
\begin{equation}\label{eq:norm-int}\begin{split}
  &X(k,k') = F_3(k,k') + \frac{1}{4}r(k)r(k') \times \\
  &\times \sum_{s,s'=\pm 1} e^{i(s\alpha(k)-s'\alpha(k'))}  
  \int_1^{\infty}\rd v \,\overline{\ub{e}_{s'k'}(v)}\,\ub{e}_{sk}(v)
\end{split}\end{equation}
Knowing the form of $\ub{e}_{\pm k}$ and the relation \cite{tables}
\begin{equation}
  \int_{\re} \rd x \, \theta(x) \, e^{ikx} 
  = \frac{1}{2}\left( \delta(k) - \frac{i}{\pi k} \right)
\end{equation}
we can evaluate the integrals in \eqref{eq:norm-int}. The result is
\begin{equation}\begin{split}
  &X(k,k') = F_4(k,k') + \frac{1}{8}r(k)r(k') \times \\
  &\times \sum_{s,s'=\pm 1} e^{i(s\alpha(k)-s'\alpha(k'))} \delta(sk-s'k') \ ,
\end{split}\end{equation}
where $F_4$ is again a well defined function.

Taking into account, that $k,k'\in\re^+$, thus the test functions integrated with the 
distribution have support only at positive $k'$ we get the relation
\begin{subequations}\label{eq:nd-equal}\begin{align}
  \delta(k-k') &= X(k,k') = F_4(k,k') \tag{\ref{eq:nd-equal}} \\
  &+ \frac{1}{4}\cos(\alpha(k)-\alpha(k'))r(k)r(k')\delta(k-k') \ , \notag
\end{align}\end{subequations}
which can satisfied in the distributional sense only if $r(k)=2$. In consequence, 
given a WDW limit $\ub{\psi}_k$ of the LQC basis function $e_k$ as defined in 
\eqref{eq:WDW-lim-def}, the following holds
\begin{equation}\label{eq:norm-rel}
  \| \psi_k \| = 2\sqrt{2} \| e_k \| .
\end{equation}

\subsection{Applicability of the stationary phase method}
\label{app:Ftr}

In Sec.~\ref{sec:triangle-wdw} the stationary phase method was applied to approximate 
the integral \eqref{eq:tN-def} determining the proportionality factor $\tilde{N}$.
However, as the integrated function is singular in $b=0$ and the integration is 
performed over the real line, the question whether 
the method can be applied there, is nontrivial. Here we show, that the contribution 
from the neighborhood of the singularity can be neglected and the method selected 
can be in fact applied.

Consider now the integral in \eqref{eq:tN-def}. It can be split onto two parts.
\begin{subequations}\label{eq:N-split}\begin{align}
  \begin{split}
    L(k) &:= \int_{\re}\rd y\, e^{-ik(\ln|y|-y)} \\
    &\hphantom{:}= L^+(k) + L^-(k) + \tilde{L}(k) ,
  \end{split} \\
  L^{\pm}(k) &:= \int_{\re^{\pm}}\rd y\, \xi(y)e^{-ik(\ln|y|-y)} , \\
  \tilde{L}(k) &:= \int_{\re}\rd y\, (1-\xi(y))e^{-ik(\ln|y|-y)},
\end{align}\end{subequations}
where $\xi(y)$ is a smooth function equal to
\begin{equation}
  \xi(y) := \begin{cases}
              0, & |y-1|\le\rho\\
              *, & \rho<|y-1|< 2\rho\\
              1, & |y-1|\ge 2\rho 
            \end{cases}
\end{equation}
for some chosen small $\rho$. Stationary phase method can be safely applied 
to the term $\tilde{L}(k)$. On the other hand one can show, that the remaining
terms are bounded by the function $O(k^{-1})$. Indeed, integrating $L^{+}(k)$ by parts  we get 
\cite{part-note} 
\begin{subequations}\label{eq:int-parts}\begin{align}
  L^+(k) &= \frac{i\epsilon \xi(\epsilon)e^{-ik(\ln(\epsilon)-\epsilon)}}{k(1 -\epsilon)}\Big|_{\epsilon=0}\notag\\
     &\phantom{=}- \int_0^\infty\rd y\, \xi'(y)\frac{iye^{-ik(\ln(y)-y)}}{k(1-y)}\notag\\
     &\hphantom{=}- \int_0^\infty\rd y\, \frac{i}{k}\xi(y)\frac{\partial}{\partial y}
     \left(\frac{y}{1-y}\right)e^{-ik(\ln(y)-y)}, \tag{\ref{eq:int-parts}}
\end{align}\end{subequations}
where the integral in the $3$rd line (denoted further as $I_2(k)$) 
can be rewritten as
\begin{equation}\label{eq:part-int-form}
  I_2(k) = \int_\epsilon^\infty\rd y\, \xi(y)\frac{i}{k(1-y)^2}e^{-ik(\ln(y)-y)} .
\end{equation}
The form of both the integrals above and the function $\xi(y)$ allows us to estimate 
$L^+(k)$ as
\begin{equation}
  |L^+(k)| \leq \frac{\xi_1}{k} + \frac{\xi_2}{k} 
\end{equation}
where $\xi_i$ are some constants common for all $k$. In consequence we obtain 
an estimate for the integral which is of the order $O(1/k)$. The same technique 
can be applied to $L^-(k)$, giving the bound of the same order.

The similar estimate can be derived for the integral \eqref{eq:N'int} determining $\tilde{N}'$
by introducing the splitting analogous to \eqref{eq:N-split}. Then the term $L'^+(k)$ 
--an analog of $L^+(k)$-- can be reexpressed as
\begin{equation}\label{eq:diff-int-parts}\begin{split}
  L'^+(k) &= \int_0^{\infty}\rd y\, \xi(y)\ln(y) e^{-ik(\ln(y)-y)}  \\
  &=\frac{i\xi(\epsilon)\epsilon\ln(\epsilon)}{k(1-\epsilon)}
     e^{-ik(\ln(\epsilon)-\epsilon)}\Big|_{\epsilon=0}   \\
   &\hphantom{=}-\int_0^{\infty}\rd y\, \frac{i}{k}\xi(y)
    \frac{\partial}{\partial y}\left(\frac{y\ln y}{1-y}\right) e^{-ik(\ln(y)-y)}  \\
   &\hphantom{=}-\int_0^{\infty}\rd y\, \frac{i}{k}\xi'(y)\frac{y\ln y}{1-y}
    e^{-ik(\ln(y)-y)} , 
\end{split}\end{equation}
where the first righthand side integral (denoted as $I_1(k)$) equals
\begin{equation}\begin{split}
  I_1(k)&=\int_0^{\infty}\rd y\, \xi(y)\frac{i}{k}\frac{1-y+\ln(y)}{(1-y)^2} e^{-ik(\ln(y)-y)}\\
  &=\int_0^{\infty}\rd y\, \xi(y)\frac{i}{k}\frac{1}{1-y} e^{-ik(\ln(y)-y)}\\
  &\phantom{=}+\int_0^{\infty}\rd y\, \xi(y)\frac{i}{k}\frac{\ln(y)}{(1-y)^2} e^{-ik(\ln(y)-y)} .
\end{split}\end{equation}
Applying again the integration by parts to the first integral, we obtain the following estimate
\begin{equation}
  |I_1(k)|\leq \frac{\xi_1}{k^2}+ \frac{\xi_2}{k} .
\end{equation}
As a result, we obtain an estimate on $L'^+(k)$
\begin{equation}
  |L'^+(k)|\leq \frac{\xi_1}{k^2}+ \frac{\xi_2}{k}+ \frac{\xi_3}{k},
\end{equation}
where, as before, one can choose a common values of $\xi_i$ for all $k$. Analogously 
we arrive to the similar estimate on $L'^-(k)$ -- an analog
of $L^-(k)$.

\subsection{Relations between the dispersions and the correlations}
\label{app:triangles}

Here we briefly remind the well known relation between the correlation of two operators 
and their dispersions. That relation is general and holds for any quantum mechanical 
system. While being quite basic, it seems to be often forgotten in the present literature
in LQC, which sometimes may lead to an impression, that the correlations can grow 
uncontrollably. That relation is applied here in the derivation of the bound 
on the dispersion of the sum of the operators, which is in turn used in Sec.~\ref{sec:triangle-wdw}.

Consider two operators $\hat{A}$, $\hat{B}$ essentially self-adjoint in some dense
domain in a Hilbert space $\Hil$. Denote the expectation values of these operators
evaluated on the state $\Psi\in\Hil$ respectively as $a$ and $b$. Denote also their 
dispersions as $\sigma_A$ and $\sigma_B$. The correlation $E(A,B)$ between $\hat{A}$ 
and $\hat{B}$ is defined as the expectation value 
\begin{equation}
  E(A,B) := \langle (\hat{A}-a\id)(\hat{B}-b\id) \rangle 
  + \langle (\hat{B}-b\id)(\hat{A}-a\id) \rangle . 
\end{equation}
Applying the Schwartz inequality to the states $\ket{\Phi}:=(\hat{A}-a\id)\ket{\Psi}$ and 
$\ket{\chi}:=(\hat{B}-b\id)\ket{\Psi}$ we immediately get the bound
\begin{equation}\label{eq:correl-bound}
  \fracs{1}{2}|E(A,B)| = |\ip{\Phi}{\chi}| \leq \|\Phi\|\|\chi\| = \sigma_A \sigma_B . 
\end{equation}
The use of Schwartz inequalities allows to derive analogous relations between at least 
some of the higher order components of the Hamburger decomposition (defined for example in
\cite{boj-ham}). Thus, it might provide a useful tool for the qualitative analysis 
of the physical evolution of the states described in terms of those momenta, in 
particular allowing for some control over their dispersion growth.

Let us now consider the sum of operators $\hat{A}$ and $\hat{B}$. Its dispersion
on the state $\Psi$ by definition equals
\begin{equation}
  \sigma^2_{A+B} = \sigma^2_A + E(A,B) + \sigma^2_B .
\end{equation}
Applying the bound \eqref{eq:correl-bound} and taking the square root of the above 
expression we arrive to the inequality
\begin{equation}\label{eq:disp-sum}
  \sigma_{A+B} \leq \sigma_A + \sigma_B , 
\end{equation}
which holds for any pair of self-adjoint operators and any physical state, independently 
of whether $\hat{A}$ and $\hat{B}$ commute.

\section{Numerical aspects of the analysis}\label{app:num}

In this appendix we describe the actual numerical techniques used to investigate 
the properties of the phase shifts $\alpha(k)$ defined in \eqref{eq:WDW-lim-def}, 
which were presented in Sect.~\ref{sec:phases} and applied in 
Sect.~\ref{sec:triangle-wdw}.

The eigenfunctions $\psi_k = \tilde{r}(k) e_k$ can be calculated directly from 
\eqref{eq:eigenf} via iterative methods described in \cite{aps-imp}. Since 
\eqref{eq:eigenf} is of the $2$nd order, to determine the eigenfunction uniquely 
one in principle needs to provide its initial values at two consecutive points of 
$\lat_{\varepsilon}$. However for the prescriptions and the superselection sectors 
considered in this paper the value $\psi_k(v=8)$ is determined uniquely by 
$\psi_k(v=4)$ (see \cite{aps-imp,acs,mmo} for the details), thus the latter one 
is the single initial datum. A main disadvantage of this method is, that at the 
level of providing the initial data it is not possible to normalize the function. 
It can be done only via application of the relation \eqref{eq:norm-rel}, once $\psi_k$ 
is calculated on some large domain of $v$ and its WDW limit is found. Fortunately the phase
shifts $\alpha$ defined in \eqref{eq:WDW-lim-def} are not sensitive to the normalization, 
thus in our studies $\psi_k(4)$ (or equivalently $\tilde{r}(k)$) can be set arbitrarily. 
We fix it choosing 
\begin{equation}\label{eq:num-id}
  \psi_k(v=4) = 1 .
\end{equation}

Once $\psi_k$ is evaluated in some domain $\lat^+_{0}\cap[0,v_M]$ (where $v_M\gg|k|$),
its WDW limit can be found for example via an extrapolation to $v\to\infty$ of the 
coefficients defined in \eqref{eq:eig-coeff}. In actual simulations to find $\alpha$ 
a more stable and faster converging method was implemented. Namely the vectors 
$\vec{\chi}_k(v)$ were decomposed analogously to \eqref{eq:limit-form}, that is
\begin{equation}
  \vec{\chi}_k(v) = r(k,v) \left( \begin{array}{l} e^{i\alpha(k,v)} \\ e^{-i\alpha(k,v)} 
  \end{array} \right)
\end{equation}
and the components $\alpha(k,v)$ were calculated only at the points $v_n=4n$, where 
the signs $\sgn(\psi_k(v_n))\neq\sgn(\psi_k(v_n-4))$ with use of the observation, that the function
\begin{equation}\begin{split}
  \ub{\psi}^{n}_k(v) &:= r(k,v_n) \\
  &\times [e^{ik\alpha(k,v_n)}\ub{e}_k(v)+e^{-ik\alpha(k,v_n)}\ub{e}_{-k}(v)]
\end{split}\end{equation}
vanishes only if
\begin{equation}\label{eq:k-root}
  k\ln(v)+\alpha(k,v) = \pi/2 + m\pi ,\quad m\in\integ .
\end{equation}
The position of the roots (denoted further as $v_{o,n}=\exp(x_{o,n})$) was identified 
via the linear approximation in $x=\ln(v)$
\begin{equation}\label{eq:root-expr}\begin{split}
  x_{o,n} &:= \ln(v_{o,n}) = \ln(v_n-4) \\ 
  &\hphantom{=}- \ln\left(\frac{v_n}{v_n-4}\right) \cdot \frac{\psi_k(v_n-4)}{\psi_k(v_n)-\psi_k(v_n-4)} .
\end{split}\end{equation}
Once the set of $\alpha(k,v_{o,n})$ was determined, the limit $\alpha(k)$ was found 
via a polynomial extrapolation (Neville method) with respect to the variable 
$y = \exp(-x_o)$ at $y=0$.

Such method has been applied for example in \cite{aps-imp} to evaluate $\alpha$ as 
a basis for the analytic approximation further used for construction of the initial 
data corresponding to the symmetric states. However, while it allows to calculate 
$\alpha$ quite precisely, it cannot be directly applied to find its derivatives, 
as it can be differentiated only numerically, whence
\begin{enumerate}[(i)]
  \item the numerical errors accumulating over the steps in the computation of 
     $\alpha$ would not allow to derive sufficiently precise differential limit 
     for small displacement of $k$, and
  \item for larger displacements of $k$ one could not exclude the existence of 
     small ``large frequency'' terms which in principle, while giving negligible 
     contribution to $\alpha$ itself, could give a considerable one to its 
     derivatives.
\end{enumerate}
Here, to avoid the above problems, we use a slight modification of the method 
under discussion, namely we find $\alpha'$ by differentiating \eqref{eq:k-root} 
over $k$
\begin{equation}\label{eq:root-diff}
  [\partial_k\alpha](k,v) = - (x_o + kx'_o) , \quad x_o = \ln(v_o) ,
\end{equation}
where the values $x'_{o,n}$ of the derivative $x'_o$ are in turn determined via 
differentiating \eqref{eq:root-expr}
\begin{equation}\label{eq:a-diff}\begin{split}
  x'_{o,n} &= \ln\left(\frac{v_n}{v_n-4}\right)  \\
  &\times 
  \frac{\psi'_k(v_n)\psi_k(v_n-4)-\psi'_k(v_n-4)\psi_k(v_n)}{(\psi_k(v_n)-\psi_k(v_n-4))^2} .
\end{split}\end{equation}
The derivatives $\psi'_k(v_{o,n})$ appearing in the above expression can be in turn 
evaluated via an iterative equation resulting from a differentiation of \eqref{eq:eigenf}, 
that is
\begin{equation}\label{eq:diff1-iter}\begin{split}
  \psi'_k(v+4) = \frac{1}{f_+(v)} [ &(\omega^2 + f_o(v))\psi'_k(v) 
  + 2\frac{\omega}{k}\psi_k(v) \\ &- f_-(v)\psi'_k(v-4) ] . 
\end{split}\end{equation}
Up to the inhomogeneous term proportional to $\psi_k(v)$ that equation is a complete
analog of \eqref{eq:eigenf}, thus the initial value problem is equivalent as well, 
once $\psi_k$ itself is known. Therefore the initial condition \eqref{eq:num-id} 
specifies uniquely not only $\psi_k$ but also $\psi'_k$.

The equations \eqref{eq:num-id}, \eqref{eq:root-diff}, \eqref{eq:a-diff} and 
\eqref{eq:diff1-iter} allow one to determine $[\partial_k\alpha](k,v_{o,n})$ corresponding
to finite $v_n$.
Then the limit $[\partial_k\alpha](k)$ is found via the polynomial extrapolation 
in the exactly same way as $\alpha(k)$.

To find $[\partial^2_k\alpha](k)$ one proceeds in the same way, first determining
$[\partial^2_k\alpha](k,v_{o,n})$ analogously to $[\partial_k\alpha](k,v_{o,n})$ via the 
system of equation build by differentiation of \eqref{eq:num-id}, \eqref{eq:root-diff}, 
\eqref{eq:a-diff} and \eqref{eq:diff1-iter} and computing the final limit by the 
polynomial extrapolation.

In the actual simulations performed the sequences $\{x_{o,n}\}_i$ of the roots were 
searched for in such a way, that an $i$th root in the sequence is the maximal root 
within the range $[0,2^{-i}v_M]$. The bound $v_M$ ranged from $10^6$ to $2\cdot 10^8$.
Depending on the number of the roots actually found the order of the polynomial used
to find the final limits varied from $0$ to $2$. The range of $k$ investigated 
numerically was $[10^{-1},10^6]$.

\begin{figure*}[p]
  \begin{center}
    $(a)$\hspace{3.2in}$(b)$
  \end{center}
  \vspace{-0.4cm}
  \includegraphics[width=3.1in]{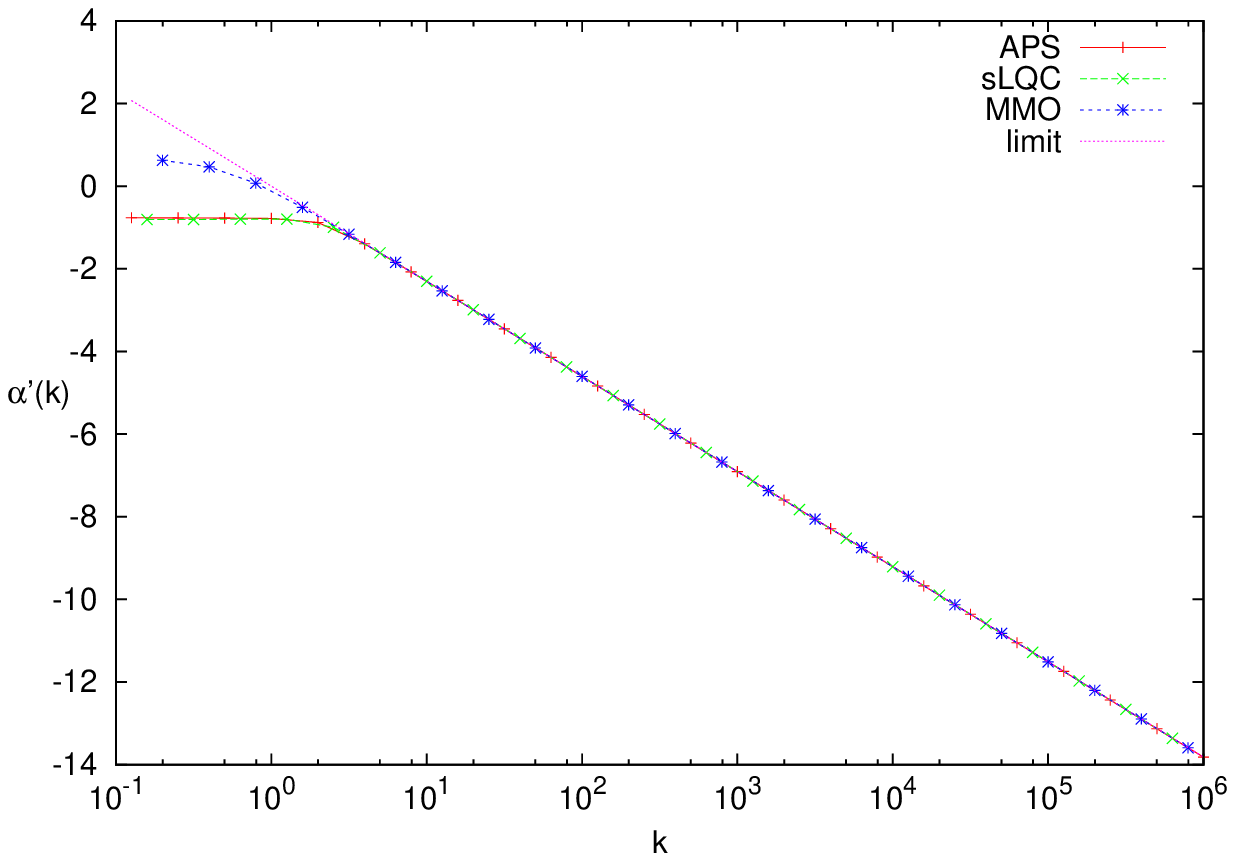}
  \includegraphics[width=3.17in]{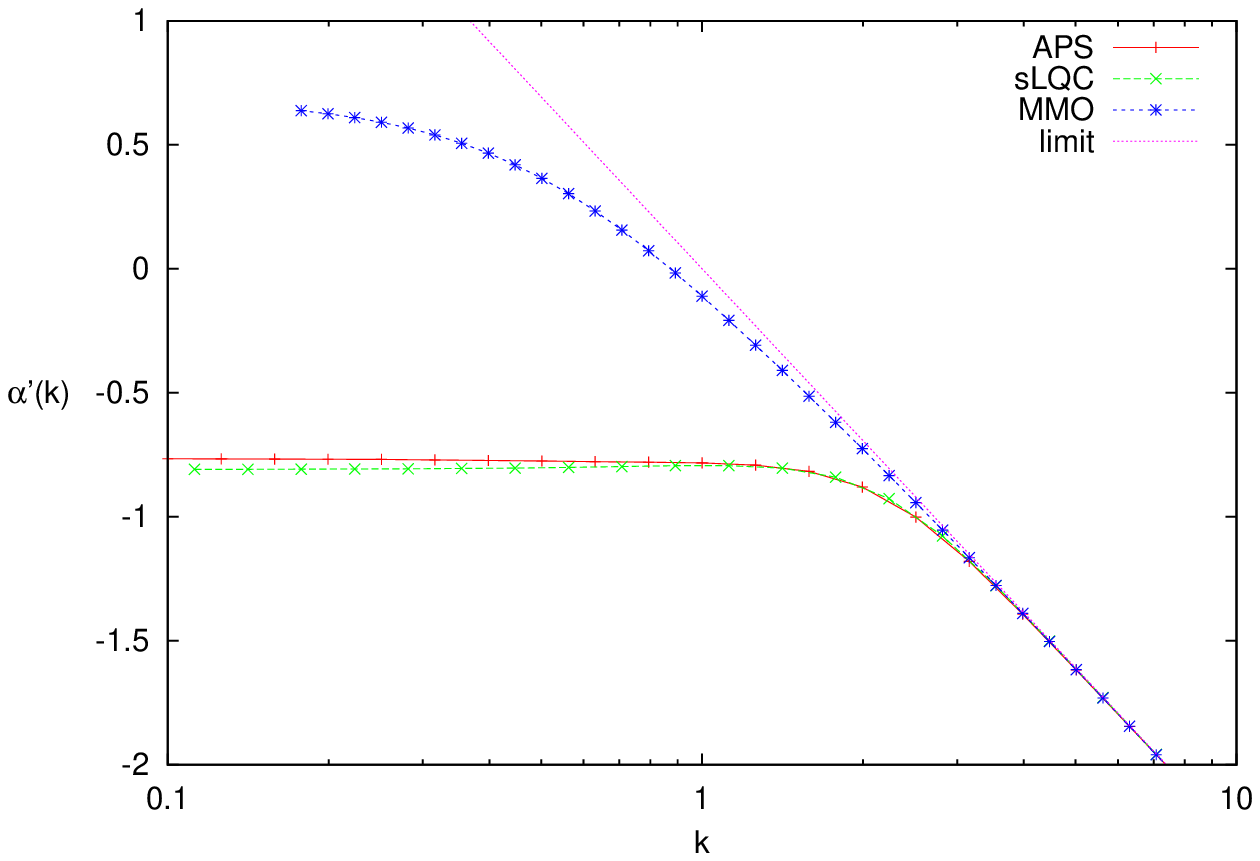}
  \caption{The behavior of $\partial_k\alpha$ for different quantization prescriptions 
    compared against its large $k$ limit $\alpha'(k)=-\ln(k)$. Figure $(a)$ shows 
    the shape of the derivative in the entire domain investigated numerically, 
    whereas $(b)$ presents the zoom for small values of $k$.}
  \label{fig:diff1}
\end{figure*}
\begin{figure*}[p]
  \begin{center}
    $(a)$\hspace{3.2in}$(b)$
  \end{center}
  \vspace{-0.4cm}
  \includegraphics[width=3.1in]{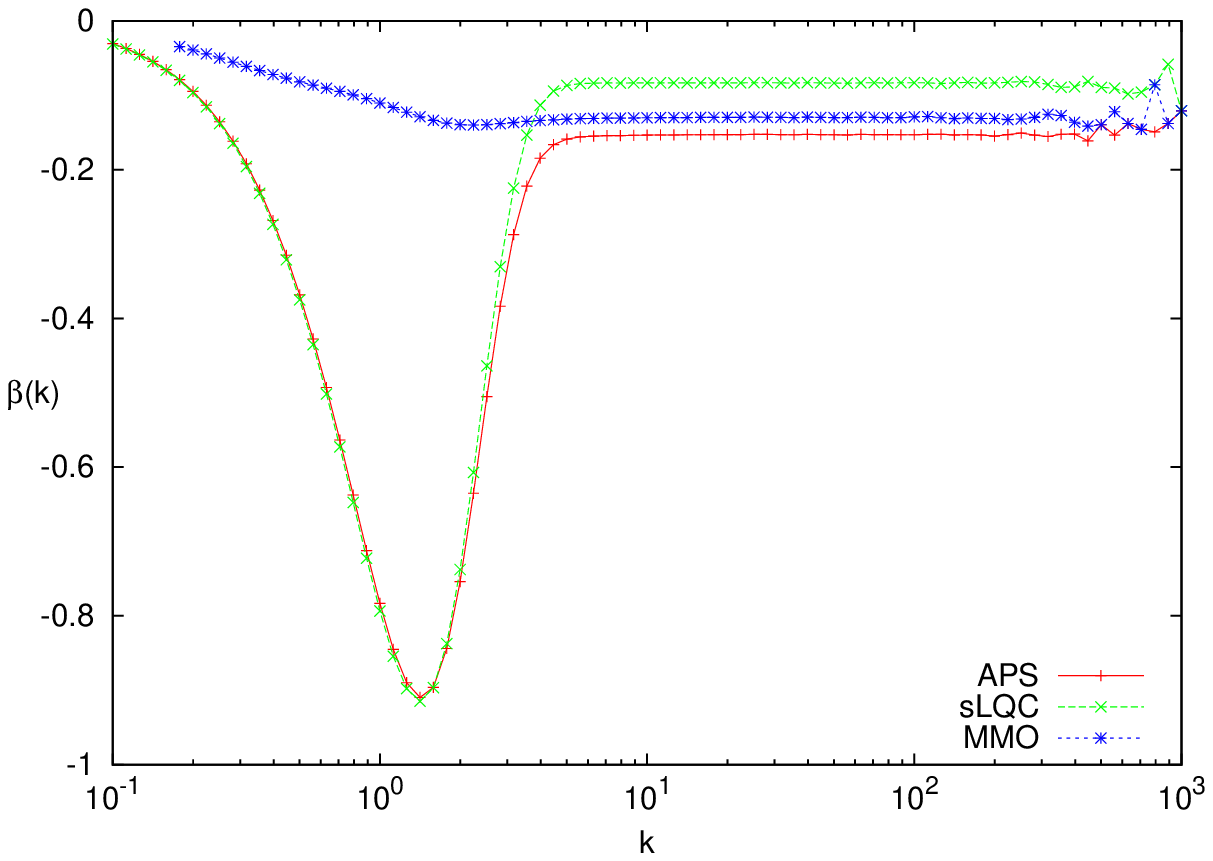}
  \includegraphics[width=3.1in]{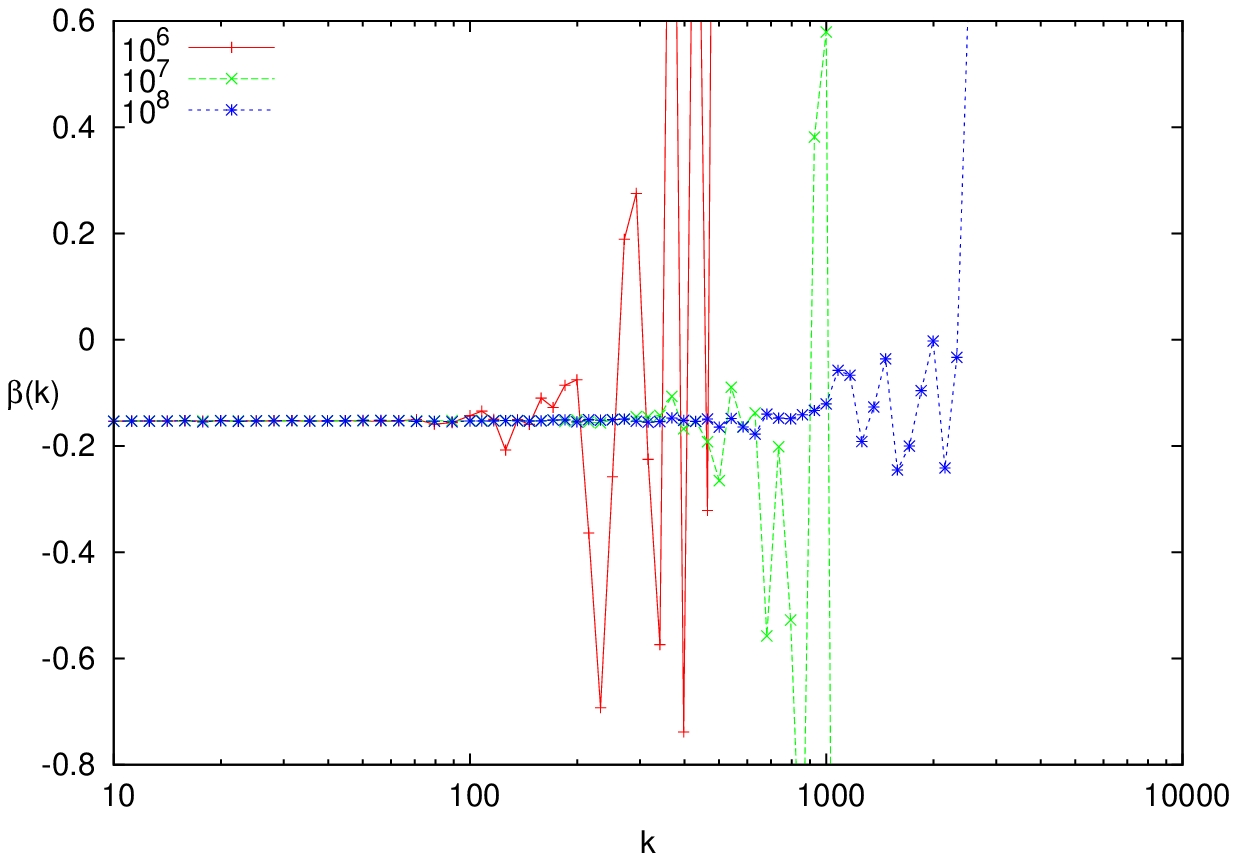}
  \caption{The values of the rescaled remnant function $\beta(k):=(\alpha'(k)+\ln(k))k^2$
    calculated numerically within the domain of reliability of the applied method. Figure
    $(a)$ shows the result for different prescriptions whereas $(b)$ presents the dependence
    of calculated $\beta$ on the choice of $v_M$ on the example of the APS prescription.
    One can see that as $v_M$ increases $\beta(k)$ converges.}
  \label{fig:diff1rem}
\end{figure*}
\begin{figure*}[p]
  \begin{center}
    $(a)$\hspace{3.2in}$(b)$
  \end{center}
  \vspace{-0.4cm}
  \includegraphics[width=3.19in]{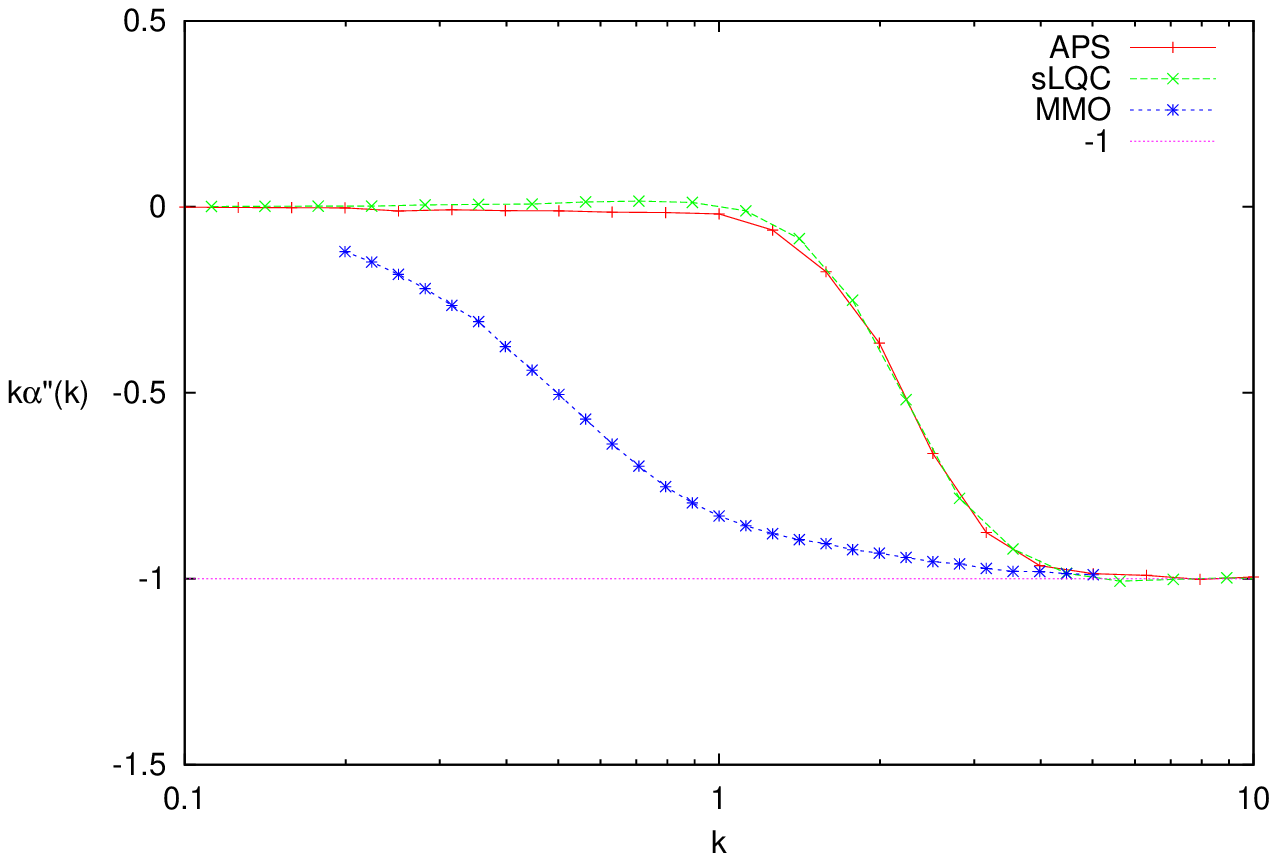}
  \includegraphics[width=3.1in]{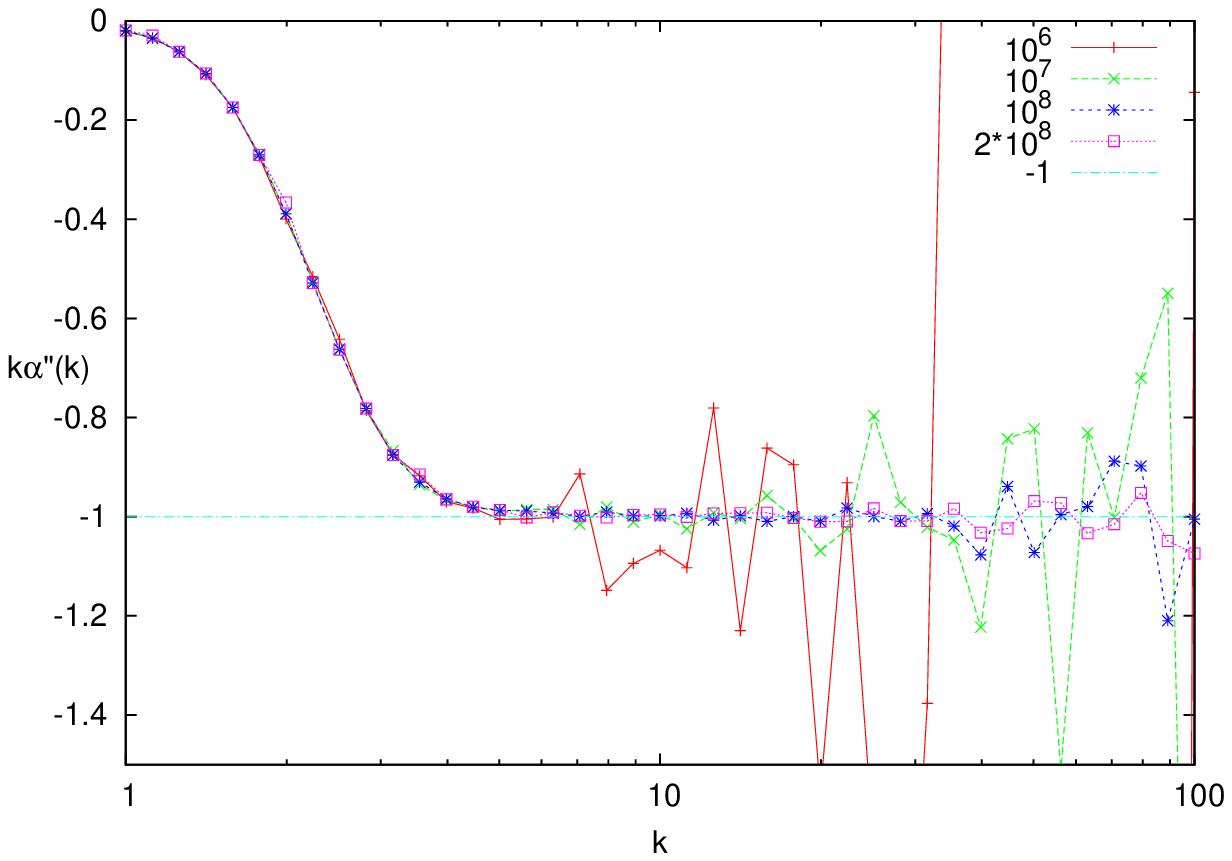}
  \caption{$(a)$ The rescaled derivative $k\partial^2_k\alpha$ for different quantization 
    prescriptions shown in the domain, where it could be determined numerically 
    in the reliable manner.
    $(b)$ The values of $k\partial^2_k\alpha$ for APS prescription evaluated on
    different sets of probing points with the upper bound ranging from $10^7$ to
    $2\cdot 10^8$. The numerical solutions converge as the bound increases.}
  \label{fig:diff2}
\end{figure*}

The results are presented in Figs.~\ref{fig:diff1} to \ref{fig:diff2}. As we can 
see in Fig.~\ref{fig:diff1} the chosen methods allow to calculate $\partial_k\alpha$ 
quite precisely within the entire investigated domain. Its precise form varies for 
small $k$ depending on the particular prescription, whereas for large $k$ the derivative
quickly approaches $-\ln(k)$. The rate of approach \eqref{eq:dalimit} is confirmed 
for $k<10^3$ (see Fig.~\ref{fig:diff1rem}a), however due to the numerical errors
the results are no longer reliable for larger $k$. Nonetheless the results converge 
also there as $v_M$ increases, which is shown in Fig.~\ref{fig:diff1rem}b. 

The $2$nd order derivative could be evaluated reliably only for $k\leq 10$, as 
it is more sensitive to the numerical errors. The result is presented in Fig.~\ref{fig:diff2}a
showing the rescaled derivative $k\partial^2_k\alpha$. From there we observe that
it quickly approaches its limiting value $k\partial^2_k\alpha \to -1$ deduced from
Fig.~\ref{fig:diff1}a, never exceeding it \cite{bound-exc}. As in the case of the 
remnant of $\partial_k\alpha$ the numerical results converge as $v_M$ increases, 
which is presented in Fig.~\ref{fig:diff2}b.


\begin{thebibliography}{99}
\bibitem{rep-diff} In comparizon to \cite{aps-imp} here we use slightly different 
    representation obtained from the one of APS via unitary mapping $\phi(v)
    \mapsto [B(v)]^{1/2}\phi(v)$.
\bibitem{ind-drop} For the simplicity we drop the index $\varepsilon$, remembering 
    that here we consider just the sector $\varepsilon=0$.
\bibitem{remnants} Here the symbols $O(f(v))$ and $o(f(v))$ are defined via requirements that
    $\limsup_{v\to\infty}O(f(v))/f(v)<\infty$ and $\lim_{v\to\infty}o(f(v))/f(v)=0$.
\bibitem{smooth-note} The successive 
    derivatives $g^{(n)}_k$ starting from $n=2$ are integrable at $\pm\infty$.
\bibitem{norm-symm} One has to remember that in both APS and sLQC prescriptions the
    symmetric eigenfunctions considered here have support on both $v>0$ and $v<0$. 
    Therefore in those cases we have in fact integrals over $[1,\infty[$ as well as 
    $]-\infty,-1]$, whereas in MMO prescription we have just one orientation and one 
    integral. On the other hand, when dealing with symmetric LQC states we should also
    symmetrize the WDW ones and the orthonormality condition should apply to the symmetrized
    WDW basis. In consequence, while the derivation is formally formulated just 
    for one orientation (MMO prescription) the final expression \eqref{eq:norm-rel} 
    remains valid also in the cases of symmetric bases (APS and sLQC prescription).
\bibitem{part-note} To be more precise, one should use here the regularization 
    $e^{-\eta x}$ which removes the upper boundary term in the integration if 
    it grows slower than any exponential function.
\bibitem{bound-exc} The points, for which $|k\partial^2_k\alpha|>1$, observed in Fig.~\ref{fig:diff2}a 
    are the result of the numerical errors, see Fig.~\ref{fig:diff2}b.
\end{thebibliography}
\end{document}